\documentclass[twocolumn,amsmath,amssymb,aps]{aastex62}
\usepackage[fleqn]{amsmath}
\usepackage{mathrsfs}
\usepackage{ulem}

\usepackage{ctable}
\usepackage{graphicx}
\usepackage{dcolumn}
\usepackage{bm}
\usepackage{makecell}
\usepackage{footmisc}

\received{}
\revised{}
\accepted{April 22, 2021}
\submitjournal{The Astrophysical Journal}

\shorttitle{Mass-radius relationships for irradiated ocean planets}
\shortauthors{Aguichine et al.}

\begin{document}
	
\title{Mass-radius relationships for irradiated ocean planets}

\author{Artyom Aguichine}
\email{artem.aguichine@lam.fr}
\affil{Aix Marseille Univ, CNRS, CNES, LAM, Marseille, France}
\author{Olivier Mousis}
\affil{Aix Marseille Univ, CNRS, CNES, LAM, Marseille, France}
\author{Magali Deleuil}
\affil{Aix Marseille Univ, CNRS, CNES, LAM, Marseille, France}
\author{Emmanuel Marcq}
\affiliation{LATMOS/IPSL, UVSQ Université Paris-Saclay, Sorbonne Université, CNRS, Guyancourt, France}

\begin{abstract}
Massive and water-rich planets should be ubiquitous in the universe. Many of those worlds are expected to be subject to important irradiation from their host star, and display supercritical water layers surrounded by extended steam atmospheres. Irradiated ocean planets with such inflated hydrospheres have been recently shown to be good candidates for matching the mass-radius distribution of sub-Neptunes. Here we describe a model that computes a realistic structure for water-rich planets by combining an interior model with an updated equation of state (EoS) for water, and an atmospheric model that takes into account radiative transfer. We find that the use of non appropriate EoSs can lead to the overestimation of the planetary radius by up to $\sim$10\%, depending on the planet size and composition. Our model has been applied to the GJ 9827 system as a test case and indicates Earth- or Venus-like interiors for planets b and c, respectively. Planet d could be an irradiated ocean planet with a water mass fraction of $\sim$$20\pm10\%$. We also provide fits for the mass-radius relationships, allowing one to directly retrieve a wide range of planetary compositions, without the requirement to run the model. Our calculations finally suggest that highly irradiated planets lost their H/He content through atmospheric loss processes, and that the leftover material led to either super-Earths or sub-Neptunes, depending on the water mass fraction.
\end{abstract}

\keywords{Exoplanets(498);Hydrosphere(770);Planetary interior(1248);Planetarytheory(1258);Exoplanet astronomy(486);Exoplanet structure(495), methods: numerical}

\section{Introduction}

Since oxygen is the third most abundant element in the protosolar nebula \citep{An89,Lo09}, this naturally makes water as the most abundant volatile compound in planetary bodies of our solar system, if one excepts the hydrogen and helium presents in the envelopes of the giant planets \citep{En08,Boc17,Gr17}.

Water-rich worlds (Europa, Titan, Enceladus, Pluto, Triton, etc) are ubiquitous in our solar system, and the building blocks of Uranus and Neptune are also supposed to be water rich \citep{Mo18}. These properties led astronomers to consider the possible existence of massive water-rich planets around other stars, i.e. the so-called ocean planets \citep{Le04}. Those planets would have grown from ice-rich building embryos formed beyond the snowline in protoplanetary disks, and would have subsequently migrated inward up to their current orbital location nearby their host star \citep{Ra18a,Ra18b}. This motivated the implementation of an H$_2$O layer to existing internal structure models, in which the liquid water had a simple prescription for the temperature profile (often isothermal), which often led to the coexistence of liquid water with high pressure ices \citep{So07,Va07,Fo07,Ze13,Ze19}. At that time, it was believed that the temperature structure had a minor impact on the radii as it is the case for telluric planets \citep{Va06,Fo07}.

However, exoplanets considered today as good candidates for being water-rich worlds are also subject to important irradiation from their host star due to their short orbital periods. For such conditions at the surface of the planet, assuming an adiabatic temperature gradient produces very shallow P(T) profiles \citep{Th16}. As a consequence, water is not in condensed phase, but rather in supercritical state in most of their hydrospheres, making ocean planets way more inflated with an adiabatic prescription compared to an isothermal one \citep{Tu20,Mo20,Ha20}.

The inflated hydrospheres of irradiated supercritical ocean planets have been recently shown to be good candidates to account for the large radii of sub-Neptunes planets \citep{Mo20}. They could also provide a possible explanation for the bimodal distribution of super-Earth and sub-Neptune populations, also known as the Fulton gap \citep{Fu17}. These physical properties, along with the availability of several sets of thermodynamic data for H$_2$O \citep{Wa11,Du06,Ma19,Jo20}, has recently motivated the modeling of the equation of state (EoS) of water in conditions relevant to planetary interiors, from 0 to a few TPa, the latter value corresponding to a Jupiter-mass planet fully made of water \citep{Ma19,Tu20,Ha20}.

For the sake of precision, mass-radius relationships of supercritical ocean planets must be calculated via the simultaneous use of atmosphere and interior structure models that are both connected at their boundaries. For example, \cite{Tu20} focused on planets of masses $0.2$--$2$ $M_\Earth$ and water contents of $0.01$--$5$ wt\%, to investigate the presence of water in the planets of the TRAPPIST-1 system. They added an irradiated steam atmosphere on top of rocky cores, using tabulated mass-radius relationships of \cite{Ze16}. These latters were computed at a 1 bar surface pressure, and might become invalid in the case of heavy H$_2$O layers (surface pressures considered up to 10 GPa). In the approach presented in \cite{Mo20}, the atmosphere model from \cite{Mar19} only considers the uppermost part of the hydrosphere up to a given pressure. The rest of the interior structure, including extreme phases of H$_2$O, is computed via an interior model \citep{Br17}, allowing to compute planets with any water content. The aim of our work is to update this model by using state of the art equations of state, and to include a better connection between the atmosphere and the interior models.

To do so, we combine the three parts of an hypothetical supercritical planet (refractory interior, condensed-fluid H$_2$O layer, and steam atmosphere) in a self-consistent framework to provide analytical descriptions of mass-radius relationships, which depend on the planetary mass, water mass fraction (WMF) and the equilibrium temperature. Such a derivation will allow estimating the WMF of irradiated ocean exoplanets from ground- or space-based mass-radius observations.

We also discuss the possible existence of these supercritical planets in light of hydrodynamic and Jeans' atmospheric escapes, and provide the mass-radius domains where escape is efficient. We finally use our model to compute the WMF of exoplanets b, c, and d of the system GJ-9827, chosen as a test case, and find that planet d could be a planet in supercritical state made of $20\pm 10\%$ of H$_2$O by mass.

Section \ref{sec:previous} reviews the model from \cite{Mo20}, presenting its main features, inputs and outputs. Section \ref{sec:improvements} details the work that has been made to update the model's EoS and make a consistent connection between the interior and the atmosphere model. Results are shown in Section \ref{sec:results} in the form of mass-radius relationships, and ternary diagrams, and a conclusion is made in Section \ref{sec:ccls}.

\section{Underlying interior and atmospheric models} \label{sec:previous}

We follow the approach of \cite{Mo20} consisting in coupling a Super-Earth interior model derived from \cite{Br17} and the atmospheric model described in \cite{Ma17,Mar19}. Here we recall the basic assumptions of these models.

\subsection{Interior Model}

Our model solves iteratively the equations describing the interior of a planet:

\begin{eqnarray}
\frac{\mathrm{d} g}{\mathrm{d} r}&=&4 \pi G \rho-\frac{2 G m}{r^{3}}, \label{eq:gauss}\\
\frac{\mathrm{d} P}{\mathrm{d} r}&=&-\rho g, \label{eq:hydrostatic}\\
\frac{\mathrm{d} T}{\mathrm{d} r}&=&-g\gamma T\frac{\mathrm{d} \rho}{\mathrm{d} P}, \label{eq:temp_grad}\\
P &=& f(\rho,T), \label{eq:solve_eos}
\end{eqnarray}

\noindent which correspond to the Gauss's theorem, hydrostatic equilibrium, adiabatic profile with use of Adams-Williamson equation, and the EoS of the considered medium, respectively. $g$, $P$, $T$ and $\rho$ are gravity, pressure, temperature and density profiles, respectively. $m$ is the mass encapsulated within the radius $r$, $G$ is the gravitational constant, and $\gamma$ is the Gr\"uneisen parameter. The Gruneïsen parameter is key to compute the thermal profile of the planet, and the literature sometimes refers to the adiabatic gradient instead, expressed as follows \citep{Ki12,Ma19,Ha20}:

\begin{eqnarray}
	\nabla_\mathrm{ad} = \left(\frac{\partial \ln T}{\partial \ln P}\right)_S = \gamma \frac{P}{\rho} \frac{1}{c^2},
\end{eqnarray}
where $S$ is the entropy, and $c$ the speed of sound.\\

The interior model can display up to five distinct layers, depending on the planet's characteristics:

\begin{itemize}
	\item a core made of metallic Fe and FeS alloy;
	\item a lower mantle made of bridgmanite and periclase;
	\item an upper mantle made of olivine and enstatite;
	\item an ice VII phase;
	\item a hydrosphere covering the whole fluid region of H$_2$O.
\end{itemize}

The Vinet EoS \citep{Vi89} with thermal Debye correction is used for all solid phases:
\begin{eqnarray}
\begin{aligned}
P\left(\rho, T\right)=& 3 K_{0}\left[\left(\frac{\rho}{\rho_{0}}\right)^{\frac{2}{3}}-\left(\frac{\rho}{\rho_{0}}\right)^{\frac{1}{3}}\right] \times \\
& \exp\left\{\frac{3}{2}\left(K_{0}^{\prime}-1\right)\left[1-\left(\frac{\rho}{\rho_{0}}\right)^{-\frac{1}{3}}\right]\right\}\\
& + \Delta P,
\end{aligned}
\end{eqnarray}

\noindent with

\begin{eqnarray}
\begin{aligned}
\Delta P = &9 \frac{\gamma \rho R}{M_\mathrm{mol} \theta^{3}} \times\\
&\left[T^{4} \int_{0}^{\frac{\theta}{T}} \frac{t^{3}}{e^{t}-1} d t-T_{0}^{4} \int_{0}^{\frac{\theta}{T_{0}}} \frac{t^{3}}{e^{t}-1} d t\right],
\end{aligned}
\end{eqnarray}

\noindent where $\theta=\theta_{0}\left(\frac{\rho}{\rho_{0}}\right)^{\gamma}$, $\gamma=\gamma_{0}\left(\frac{\rho}{\rho_{0}}\right)^{-q}$, $R$ the ideal gas constant and $M_\mathrm{mol}$ the molar mass of the considered material. All quantities with an index 0 are reference parameters obtained by fit on experimental data, given in table \ref{tab:parameters}. The EoS used by \cite{Mo20} to solve Eq. (\ref{eq:solve_eos}) is the one formulated by \cite{Du06}, valid up to 10 GPa and 2573.15 K.

All thermodynamic and compositional parameters of mineral layers are taken equal to those of Earth \citep{St05,So07,So10}, and summarized in table \ref{tab:parameters}. We refer the reader to \cite{Br17} to get all the computational details. 

\cite{Mo20} computed the Gr\"uneisen parameter for water via a bilinear interpolation in a grid generated from the python library of the IAPWS formulation\footnote{https://pypi.org/project/iapws/\#description}, computing the Gr\"uneisen parameter in the form $\gamma = f(\rho,T)$ with $\rho$ and $T$ varying in the 316--2500 kg.m$^{-3}$ and 650--10,000 K ranges, respectively. An important issue is that the density range is very limited, since this quantity can easily vary from $\sim$10 kg.m$^{-3}$ at the planetary surface to $\sim$5000  kg.m$^{-3}$ at the center of a 100\% water planet of 1 $M_{\Earth}$, implying that the computation of $\gamma$ is erroneous at the top and at the bottom of the hydrosphere. A solution for overcoming this limitation is provided in Section \ref{sec:gruneisen}.

Apart from compositional inputs, the main physical inputs of the model are the core mass fraction (CMF) $x_\mathrm{core}$ and water mass fraction (WMF) $x_{\mathrm{H}_2 \mathrm{O}}$, the mantle mass fraction is then $x_\mathrm{mantle} = 1 - x_\mathrm{core} -x_{\mathrm{H}_2 \mathrm{O}}$. Pressure and temperature profiles are integrated from outside, and require the inputs of the boundary pressure $P_\mathrm{b}$ and boundary temperature $T_\mathrm{b}$. Finally, the model also requires the input of the planet's mass $M_\mathrm{b}$ (subscript $b$ denotes the mass encapsulated within the boundary of the interior model, excluding the contribution of any potential atmosphere). Once defined, these input parameters allow for the computation of the planet's internal structure and associated boundary radius. In the case of the Earth ($x_\mathrm{core}=0.325$, $x_{\mathrm{H}_2 \mathrm{O}}=0.0005$, $M_\mathrm{b}=1$ M$_\Earth$), the model computes a radius $R_\mathrm{b}$ equal to 0.992 $R_\Earth$, which is less than 1\% of error, indicating that errors from the model are negligible compared to errors on measurements. In the following, subscript $b$ refers to quantities at the boundary between the interior model and the atmosphere model, such as bulk mass $M_\mathrm{b}$, radius $R_\mathrm{b}$, gravity $g_\mathrm{b}$, pressure $P_\mathrm{b}$ and temperature $T_\mathrm{b}$.

\begin{table*}[!ht]
	\movetabledown=5cm
	\begin{rotatetable*}
	\centering
	\caption{List of thermodynamic and compositional parameters used in the interior model.} 
	\label{tab:parameters}
	\begin{tabular}{llllllllllll}
		\hline
		\multicolumn{2}{l}{\textit{Layer}}                     & \multicolumn{2}{l}{\textit{Core}}   & \multicolumn{4}{l}{\textit{Lower mantle}}         
             & \multicolumn{4}{l}{\textit{Upper Mantle}}                  
           \\ \hline
		Phases                              &                  & \multicolumn{2}{l}{Iron rich phase} & \multicolumn{2}{l}{Perovskite} & \multicolumn{2}{l}{Periclase} & \multicolumn{2}{l}{Olivine}   & \multicolumn{2}{l}{Enstatite}         \\
		Composition (\%)                    &                  & \multicolumn{2}{l}{100}             & \multicolumn{2}{l}{79.5}       & \multicolumn{2}{l}{20.5}      & \multicolumn{2}{l}{41}        & \multicolumn{2}{l}{59}    
            \\ \cline{3-12} 
		Components                          &                  & Fe             
  & FeS              & FeSiO$_3$      & MgSiO$_3$     & FeO           & MgO           & Fe$_2$SiO$_4$ & Mg$_2$SiO$_4$ & Fe$_2$Si$_2$O$_6$ & Mg$_2$Si$_2$O$_6$ \\
		Composition (\%)                    &                  & 87             
  & 13               & 10             & 90            & 10            & 90            & 10            & 90            & 10                & 90     
           \\ \cline{3-12} 
		Molar mass (g.mol$^{-1}$)           & $M_\mathrm{mol}$ & 55.8457        
  & 87.9117          & 131.9294       & 100.3887      & 71.8451       & 40.3044       & 203.7745      & 140.6931      & 263.8588          & 200.7774          \\
		Reference density (kg.m$^{-3}$)     & $\rho_0$         & 8340           
  & 4900             & 5178           & 4108          & 5864          & 3584          & 4404          & 3222          & 4014              & 3215   
           \\
		Reference temperature (K)           & $T_0$            & 300            
  &                  & 300            &               & 300           &   
            & 300           &               & 300               &         
          \\
		Reference bulk modulus (GPa)        & $K_0$            & 135            
  &                  & 254.7          &               & 157           &   
            & 128           &               & 105.8             &         
          \\
		Pressure derivative of bulk modulus & $K_0^\prime$     & 6              
  &                  & 4.3            &               & 4             &   
            & 4.3           &               & 8.5               &         
          \\
		Reference Debye temperature (K)     & $\theta_0$       & 474            
  &                  & 736            &               & 936           &   
            & 757           &               & 710               &         
          \\
		Reference Grüneisen parameter       & $\gamma_0$       & 1.36      
       &                  & 2.23           &               & 1.45         
 &               & 1.11          &               & 1.009             &    
               \\
		Adiabatic power exponent            & $q$              & 0.91           
  &                  & 1.83           &               & 3             &   
            & 0.54          &               & 1                 &         
          \\ \hline
	\end{tabular}
	\tablecomments{Thermodynamic data are summarized in Table 1 of \cite{So07}, and compositional data are the results of model calibration for Earth 
by \cite{St05,So07}.}
	\end{rotatetable*}
\end{table*}

\subsection{Atmospheric model} \label{sec:atmo}

\begin{figure*}[ht!]
	\resizebox{\hsize}{!}{\includegraphics[angle=0,width=5cm]{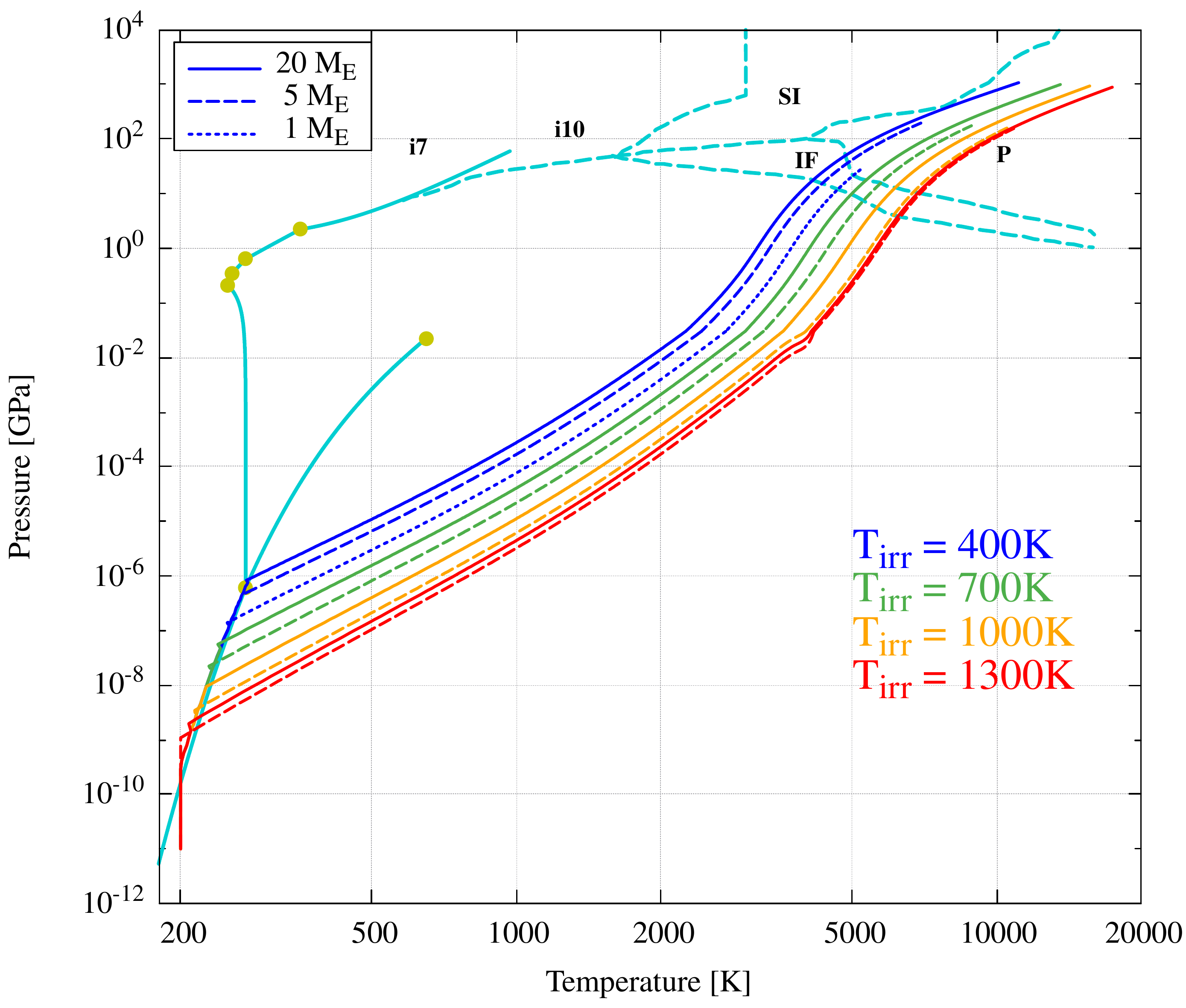}}
	\caption{$(P,T)$ profiles for 100\% H$_2$O planets of masses $M_\mathrm{p}=$1--20 $M_\Earth$, and irradiation temperatures $T_\mathrm{irr}=$400--1300 K with the Ma19+ parametrization (see section \ref{sec:gruneisen}). Cases corresponding to smallest masses and highest temperatures are not shown, as their surface gravities are below the limit fixed in Sec. \ref{sec:connect}. Phase transitions of H$_2$O are taken from \cite{Wa11} at low temperatures (solid turquoise lines) and from \cite{Ne11} at high temperatures (dashed turquoise lines), with labels IF (ionic fluid), SI (super ionic), P (plasma), and iN for the ice N.}
	\label{fig:profiles}
\end{figure*}

The atmospheric model generates the properties of a 1D spherical atmosphere of H$_2$O by integrating the thermodynamic profiles bottom to top. The model takes as inputs the planet's mass and radius, as well as the thermodynamic conditions at its bottom. We choose to connect the atmospheric model with the interior model at a pressure $P$ = $P_\mathrm{b}=300$ bar (slightly above the critical pressure $P_\mathrm{crit}=220.67$ bar) and at a temperature $T$ = $T_\mathrm{b}$. $(P,T,\rho)$ profiles are then integrated upward via the prescription from \cite{Ka88} in the case of an adiabat at hydrostatic equilibrium. Once the temperature reaches the top temperature of the atmospheric layer, here set to $T_\mathrm{top}=200$ K, an isothermal radiative mesosphere at $T=T_\mathrm{top}$ is assumed. Figure \ref{fig:profiles} shows several $(P,T)$ profiles representing the whole hydrospheres of planets (under Ma19+ parametrization, see section \ref{sec:gruneisen}) for masses and irradiation temperatures in the range 1--20 $M_\Earth$ and  $T_\mathrm{irr}=$400-1300 K (see Eq. (\ref{eq:teq})), respectively.

The atmosphere transition radius is controlled by the altitude of the top of the H$_2$O clouds, corresponding to the top of the moist convective layer, assumed to be at a pressure $P_\mathrm{top}=0.1$ Pa. We choose this limit as the observable transiting radius, assuming that results are similar for cloudy and cloud-free atmospheres \citep{Tu19,Tu20}. The EOS is taken from the NBS/NRC steam tables \citep{Ha84}, implying the atmosphere is not treated as an ideal gas. The discontinuities in $(P,T)$ profiles occuring for $T_\mathrm{irr}=1300$ K are due to the limited range of these tables, but the height of this region ($P=$ 100-300 bar) is negligible compared to the thickness of the atmosphere. Increasing the $T_\mathrm{top}$ temperature will impact the final structure of the atmosphere, decreasing both the thickness of the atmosphere and the interior. Numerical tests with $T_\mathrm{top}$ varying from 200 K to $T_\mathrm{skin}=T_\mathrm{eff}/ 2^{0.25}$ decrease the final radius of the planet of at most $\sim 200$ km for the cases considered in this study. It corresponds to a difference of $2\%$ in radius at most, but this difference is mainly below $1\%$.

Shortwave and thermal fluxes are then computed using 4-stream approximation. Gaseous (line and continuum) absorptions are computed using the $k$-correlated method on 38 spectral bands in the thermal infrared, and 36 in the visible domain. Absorption coefficients are exactly the same as those in \cite{Le13} and \cite{Tu19} which includes several databases, specificaly designed for H$_2$O-dominated atmospheres. Rayleigh opacity is also included. This method computes the total outgoing longwave radiation (OLR, in W.m$^{-2}$) of the planet that gives the temperature that the planet would have if it was a blackbody:

\begin{eqnarray}
	T_\mathrm{p} = \left(\frac{\mathrm{OLR}}{\sigma_\mathrm{sb}}\right)^{1/4}, \label{eq:tp}
\end{eqnarray}

\noindent with $\sigma_\mathrm{sb}$ the Stefan-Boltzmann constant. In order to quantify the irradiation of the planet by its host star, we define the irradiance temperature

\begin{eqnarray}
	T_\mathrm{irr} = T_\mathrm{eff} \sqrt{\frac{R_\star}{2a}}, \label{eq:tirr_obs}
\end{eqnarray}

\noindent where $T_\mathrm{eff}$ and $R_\star$ are the host star effective temperature and radius, respectively, and $a$ is the semi-major axis of the planet. The atmospheric model computes the Bond albedo from the atmosphere's reflectance \citep[][using the method presented in]{Pl19}  assuming a G-type star linking both temperatures:

\begin{eqnarray}
	T_\mathrm{irr} = \left(\frac{\mathrm{OLR}}{(1-A)\sigma_\mathrm{sb}}\right)^{1/4} = \frac{T_\mathrm{p}}{(1-A)^{1/4}}. \label{eq:teq}
\end{eqnarray}

The literature often approximates $T_\mathrm{irr}$ to the equilibrium temperature $T_\mathrm{eq}$, which is the temperature the planet would have for an albedo $A=0$ (all the incoming heat is absorbed and re-emitted by the planet). Since it is the observable quantity, our results will be presented in term of $T_\mathrm{irr}$. Equation \ref{eq:teq} assumes that the planet is in radiative equilibrium with its host star. Any heating source in the planet interior would add an additional term in the radiative equilibrium of the planet with its host star, increasing the effective temperature of the planet for the same received irradiation \citep{Ne11}. In this work, we model the structure of planets that have either no interior heating source, or that had time to cool off.

For a given planet mass, boundary radius and irradiation temperature, the atmosphere thickness and boundary temperature are retrieved from the atmospheric model. The latter is then used to compute the interior structure, and the former is taken into account to compute the total (transiting) radius.

\section{Model update} \label{sec:improvements}

This section presents the improvements made on the existing model to push further its physical limitations. Since we are interested in planets with substantial amounts of water, we define a specfic CMF, which is only related to the mass budget of the rocky part:

\begin{eqnarray}
x_\mathrm{core}^\prime = \frac{x_\mathrm{core}}{1-x_{\mathrm{H}_2 \mathrm{O}}}.
\end{eqnarray}

\noindent where $x_\mathrm{core}$ is  the ``true'' CMF. $x_\mathrm{core}^\prime$ will be used to compare planets that have different WMF, but with similar refractory contents. For example, $x_\mathrm{core}^\prime=0.325$ corresponds to an Earth-like CMF, regardless the amount of water present in the planet.

\subsection{Used EoSs}

The choice of the EoS is critical, as it strongly impacts the estimate of the mass-radius relationships. Three EoS are then considered in this study:

\begin{itemize}
	\item EoS from the latest revision of the IAPWS-95 formulation from \cite{Wa02}\footnote{http://iapws.org} (hereafter WP02). This reference EoS gives an analytical expression of the specific Helmholtz free energy $f(\rho,T)$. Any thermodynamic quantity (pressure, heat capacity, internal energy, entropy etc.) can be computed by taking the right derivative of $f$, and those quantities have analytical expressions.
	\item EoS from \cite{Du06} (hereafter DZ06). This EoS is corrected around the critical point, and gives an analytical expression for pressure as function of density and temperature $P(\rho,T)$.
	\item EoS from \cite{Ma19} (hereafter Ma19). This formulation was developed for planetary interiors by extending the IAPWS-95 EoS with ingredients from statistical physics allowing transition to plasma and superionic states. The authors created a fortran implementation\footnote{http://cdsarc.u-strasbg.fr/viz-bin/qcat?J/A+A/621/A128} that computes pressure, specific Helmholtz free energy, specific internal energy and specific heat capacity for a given couple $(\rho,T)$.
\end{itemize}

\begin{figure*}[ht!]
	\resizebox{\hsize}{!}{\includegraphics[angle=0,width=5cm]{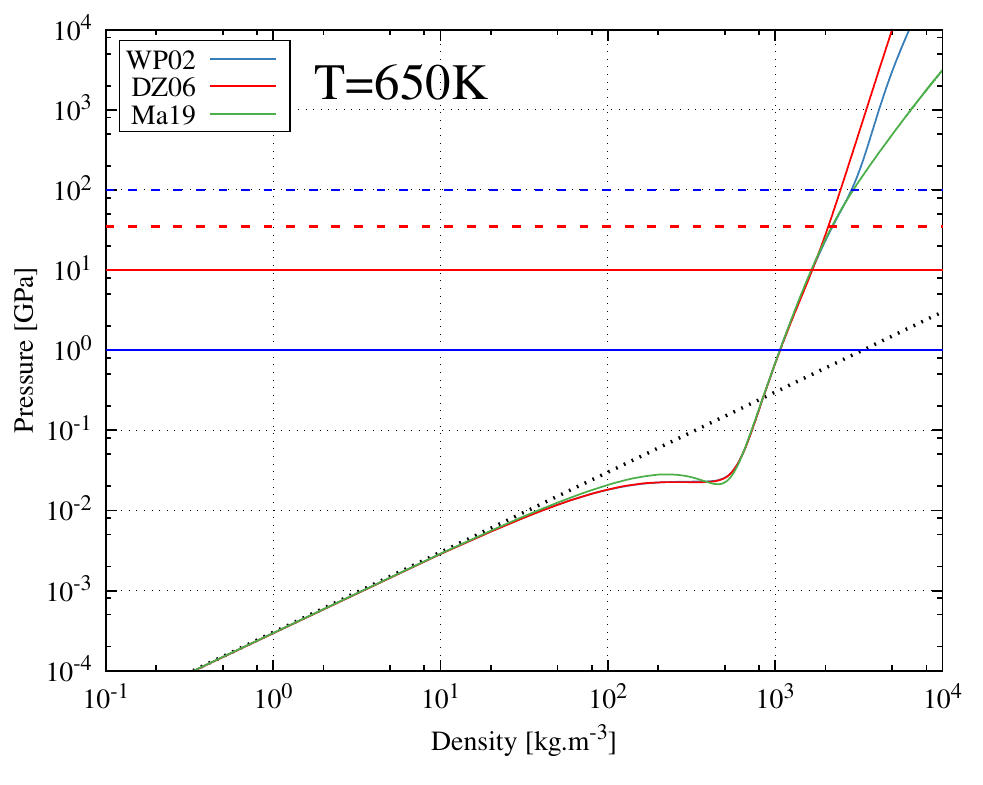}\includegraphics[angle=0,width=5cm]{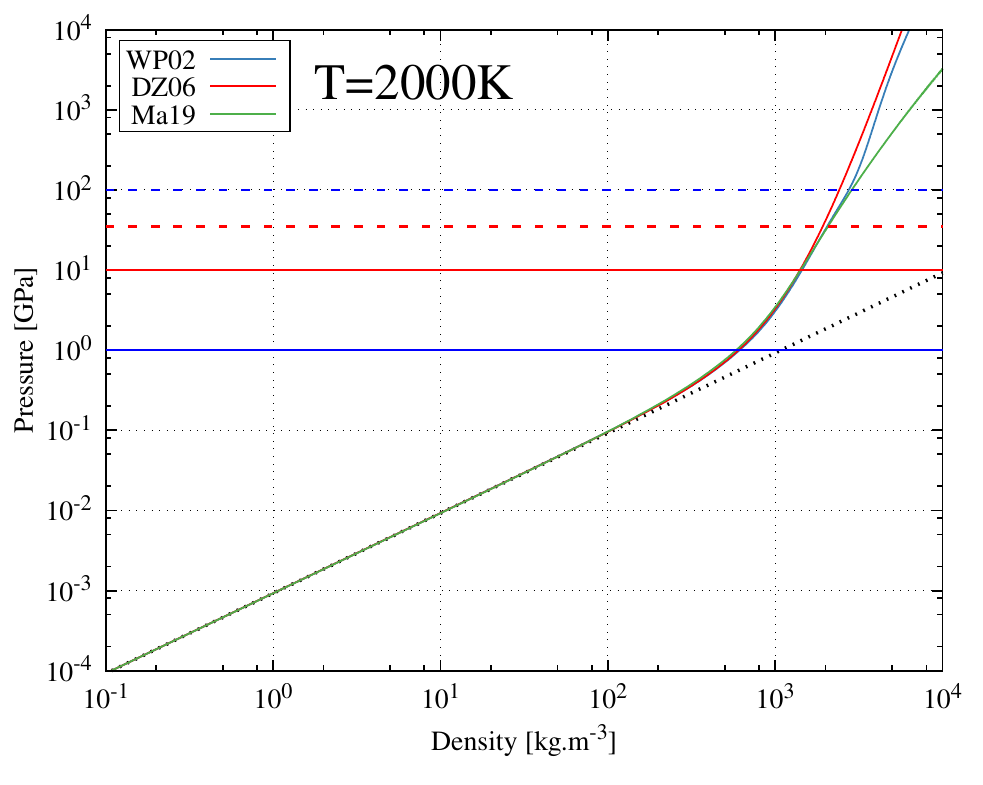} }
	\caption{Pressure as function of density calculated with WP02 (blue), DZ06 (red) and Ma19 (green) in the cases of two different temperatures. The solid horizontal lines indicate the range of validity for WP02 and DZ06, and the dashed horizontal lines give the extended range. The black dotted line corresponds to the ideal gas law for water steam.}
	\label{fig:comp_eos}
\end{figure*}

The validity ranges of the different EoSs, which rely on the availability of experimental data, are given in Table \ref{tab:range}. Extended ranges proposed by \cite{Wa02} and \cite{Du06} are also indicated because the mathematical expressions of their EoSs allow for extrapolations beyond the corresponding validity ranges. However, they become invalid when phase transition occurs (e.g. dissociation of water). Other EoSs exist in the literature, covering various regions of the phase diagram of water, or being used for specific purposes. Our choice of EoSs among others is discussed in Sec. \ref{sec:ccls}.

\begin{table}
	\centering
	\caption{Validity ranges of the different EoSs.} 
	\label{tab:range}
	\begin{tabular}{lll}
		\tablewidth{0pt}
		\hline
		\hline
		EoS 	& Valid 		& Extended 	\\	
		\hline
		WP02 & $P<1$ GPa & $P<100$ GPa   \\ 	
		&$T<1~273$ K & $T<5~000$ K \\
		DZ06  	& $P<10$ GPa & $P<35$ GPa\footnote{Limit given by \cite{Du96}.\label{limdz}}  \\
		&$T<2~573.15$ K& $T<2~800$ K\footref{limdz} 	 \\
		Ma19   	&$\rho<100\times 10^3$ kg.m$^{-3}$	&  not specified \\
		&$T<100~000$ K	&  not specified  \\
		\hline
	\end{tabular}
\end{table}

Figure \ref{fig:comp_eos} shows the $P(\rho)$ profiles derived from the considered EoSs at different temperatures. All EoSs present minor differences in their validity range, regardless the considered temperature. \cite{Ma19} find that the WP02 overestimates the pressure beyond its extended range. For a given pressure in a planet's interior, this would underestimate the density, and then overestimate the total radius of the planet. A more pronounced deviation is visible for DZ06 above its validity range. Around the critical point ($\rho\sim350$ kg.m$^{-3}$, mostly visible at 650 K), WP02 is closer to DZ06, compared to Ma19, as expected. In the low density limit, all EoSs behave following the ideal gas law $P\propto \rho T$, which has a characteristic slope of 1 in log-log scale.

\subsection{Gr\"uneisen parameter for fluids} \label{sec:gruneisen}

The Gr\"uneisen parameter $\gamma$, already introduced in Eq. (\ref{eq:temp_grad}), has many definitions. For solids, it gives the rate of change in phonon frequencies $\omega_i$ relative to a change in volume $V$ \citep{Gr12}:

\begin{eqnarray}
\gamma_i = -\left(\frac{\partial \ln \omega_i}{\partial \ln V}\right)_T.
\end{eqnarray}

By averaging over all lattice frequencies, it is possible to obtain a thermodynamic definition (using the internal energy $U$ and entropy $S$) of the Gr\"uneisen parameter \citep{Ar84} via the following expression:

\begin{eqnarray}
\gamma = V \left(\frac{\partial P}{\partial U}\right)_V = \frac{V}{C_V} \left(\frac{\partial P}{\partial T}\right)_V = \frac{\rho}{T} \left(\frac{\partial T}{\partial \rho}\right)_S. \label{eq:gamma-therm-id}
\end{eqnarray}
$\gamma$ relates a pressure (or density) variation to a temperature change. Although initialy defined for solids, the meaning of $\gamma$ holds for fluids. In planetary interiors, adiabatic heat exchange is mostly driven by convective heat transfer \citep{St19}. At planetary scales, the Gr\"uneisen parameter can thus be used for both solids and fluids. From identities in Eq. (\ref{eq:gamma-therm-id}), $\gamma$ can be expressed using other thermodynamic constants such as the thermal expansion coefficient $\alpha$, the isothermal bulk modulus $K_T$, and the specific isochoric heat capacity $c_V$
\begin{eqnarray}
\gamma = \frac{\alpha K_T }{\rho c_V}. \label{eq:gruneisen-expression}
\end{eqnarray}

\noindent  $\gamma$ is assumed to be temperature-independent in solid phase, and its value is fitted from experimental data, taking into account small density variations. In this study, we use the Helmholtz free energy $F$ given in \cite{Wa02} and \cite{Ma19}. In the IAPWS95 release, the specific Helmholtz free energy $f$ in its dimensionless form $\phi$ is divided into its ideal part (superscript $\circ$) and a residual (superscript ``r") via the following expression:

\begin{eqnarray}
\frac{f(\rho,T)}{RT} = \phi(\delta,\tau) = \phi^\circ(\delta,\tau)+\phi^\mathrm{r}(\delta,\tau),
\end{eqnarray}

\noindent with $\delta=\rho/\rho_c$ and $\tau=T_c/T$, $\rho_c$ and $T_c$ being the supercritical density and temperature, respectively. After defining the derivatives of the ideal and residual part:

\begin{eqnarray}
\phi_{m n}^{\circ}=\frac{\partial^{m+n} \phi^{\circ}(\tau, \delta)}{\partial \tau^{m} \partial \delta^{n}}, \\
\phi_{m n}^{\mathrm{r}}=\frac{\partial^{m+n} \phi^{\mathrm{r}}(\tau, \delta)}{\partial \tau^{m} \partial \delta^{n}},
\end{eqnarray}

\noindent where integers $m$ and $n$ define the order of the derivative with respect to $\tau$ and $\delta$, respectively. From these expressions, 
one can derive:

\begin{eqnarray}
\gamma_- = -\frac{1+\delta \phi^\mathrm{r}_{01}-\delta \tau \phi^\mathrm{r}_{11}}{\tau^2 \left(\phi^\circ_{20}+\phi^\mathrm{r}_{20}\right)}, \label{eq:grun-iapws}
\end{eqnarray}

\noindent where $\gamma_-$ is the formulation of the Gr\"uneisen parameter computed following the approach of  \cite{Wa02}.

The fortran implementation of \cite{Ma19} computes $F(\rho,T)$, along with other useful quantities such as $\chi_T = \left(\frac{\partial \ln P}{\partial \ln T}\right)_V$ and the specific isochoric heat capacity $c_V$. In this case, the Gr\"uneisen parameter is expressed as:
\begin{eqnarray}
\gamma_+ = \frac{P(\rho,T) \chi_T (\rho,T)}{\rho c_V T}, \label{eq:grun-mazevet}
\end{eqnarray}

\noindent where $\gamma_+$ is the formulation of the Gr\"uneisen parameter derived from the quantities calculated via the approach of  \cite{Ma19}.

In the case of an ideal gas, one can derive the theoretical value $\gamma=\frac{2}{l}$, where $l$ is the number of degrees of freedom for a given molecule. For H$_2$O,  $\gamma \simeq \frac{1}{3}$, since $l=6$ (3 rotational and 3 vibrational degrees of freedom).

The Gr\"uneisen parameter is crucial to compute the adiabatic temperature gradient inside a planet's interior.  However, because temperature has low impact on EoSs used in solid phase, it is possible to assume isothermal layers in interior models when thermodynamic data are lacking, and generate internal structures close to reality \citep{Ze19}. In the case of fluids (here H$_2$O), temperature rises sharply with depth. This strongly impacts the EoS and leads to different phase changes that are not visible in the case of isothermal profiles.

Each computation for the interior model can be performed by using any of the three EoS (WP02, DZ06, Ma19) to solve Eq. (\ref{eq:solve_eos}), and WP02 or Ma19 EoS to solve Eq. (\ref{eq:temp_grad}) (i.e. computing $\gamma$ with EoS WP02 or Ma19). In the following, we will use the name of the EoS used to solve Eq. (\ref{eq:solve_eos}), and add + or - depending on the EoS used to compute the Gr\"uneisen parameter, $\gamma_+$ (Ma19) or $\gamma_-$ (WP02) respectively. For example, Ma19- indicates that the Ma19 EoS was used to solve Eq. (\ref{eq:solve_eos}), and that the WP02 approach was used to solve Eq. (\ref{eq:temp_grad}).

\begin{figure}[ht!]
	\resizebox{\hsize}{!}{\includegraphics[angle=0,width=5cm]{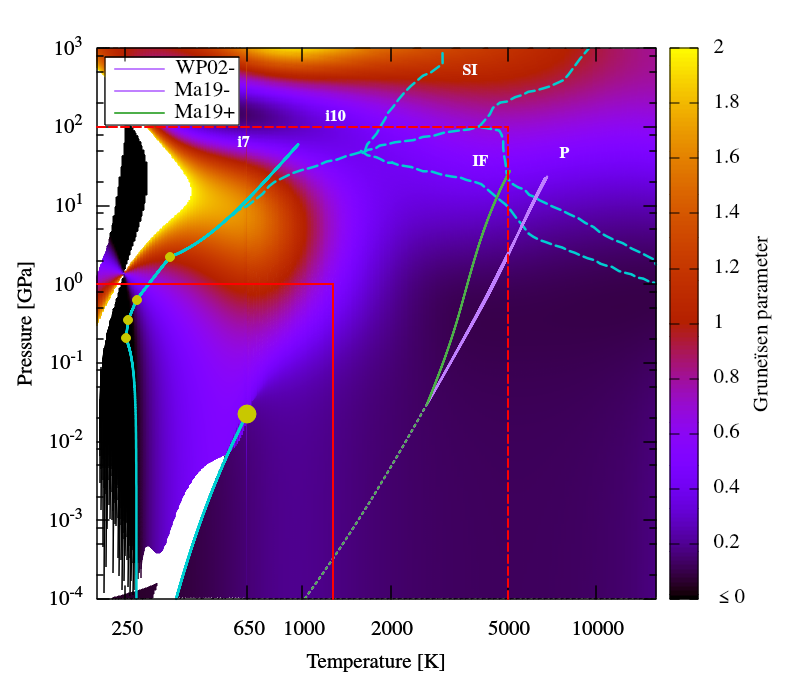}}
	\resizebox{\hsize}{!}{\includegraphics[angle=0,width=5cm]{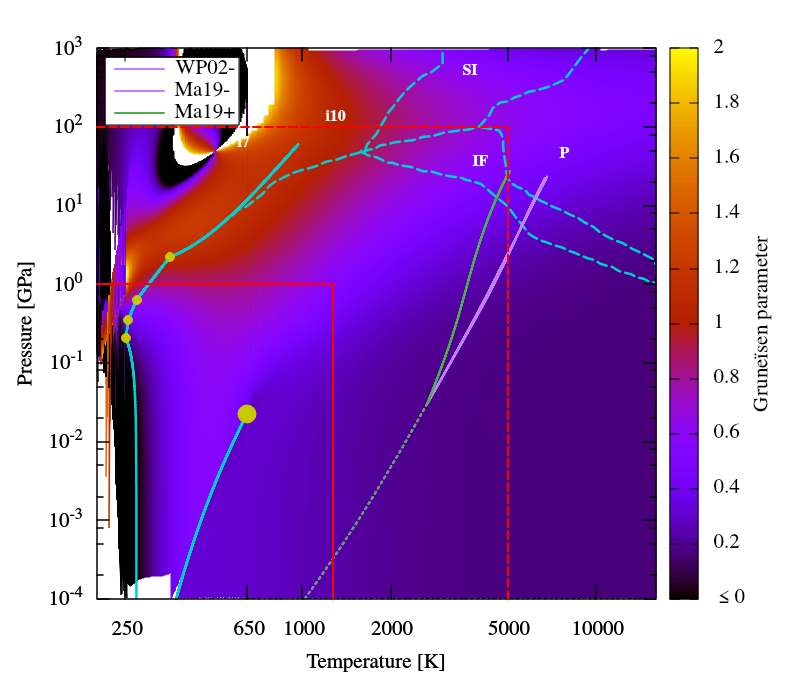}}
	\caption{Color maps showing $\gamma_+$ (top panel) and $\gamma_-$ (bottom panel) in the H$_2$O phase diagram. The phase diagram of water is  identical to the one shown in Figure \ref{fig:profiles}. Ma19+, Ma19- and WP02- are the $(P,T)$ profiles defined in the case of a 1 M$_\Earth$ planet fully made of H$_2$O, with the atmosphere part shown with short dashes. Ma19- and WP02- interiors are almost indistinguishable, hence represented by the same color, and all atmospheric profiles are identical, although the  gravity at the boundary is different for each case.}
	\label{fig:gruneisen_val}
\end{figure}

Figure \ref{fig:gruneisen_val} shows the values of $\gamma_+$ and $\gamma_-$ in the H$_2$O phase diagram. Since $\gamma$ is integrated to obtain the temperature gradient, a small difference leads to different paths in the $(P,T)$ plane. The indiscernability between the WP02- and the Ma19- profiles shows that the internal structure (and thus mass-radius relationships) is more impacted by the temperature profile than the difference in the EoS in the case of a water layer. The difference in temperature between Ma19+ and Ma19-/WP02- profiles is as high as $\sim$2000 K, which also results in a difference of $\sim 200$ kg.m$^{-3}$ in density at the center of the planet.

\subsection{Connection between interior and atmospheric models} \label{sec:connect}

Atmospheric properties (OLR, albedo, mass and thickness) are all quantities that evolve smoothly. To enable a smooth connection between the two models, we implemented a trilinear interpolation module that can estimate atmospheric properties for a planet whose physical parameters $g_\mathrm{b}$, $M_\mathrm{b}$, and  $T_\mathrm{b}$ are in the 3--30 m.s$^{-2}$,  0.2--20 $M_\Earth$, and 750--4500 K ranges, respectively. This allows us to correct the slight deviations from nods of the grid, and trilinear interpolation ensures that properties computed at a nod are exactly those at the nod, which would not be the case if a polynomial fit was performed on data. Details of the connection between the two models are given in Appendix \ref{sec:connection}.

Figure \ref{fig:teq_tp} shows $T_\mathrm{irr}$ as a function of $T_\mathrm{p}$ for a set of fixed $g_\mathrm{b}$ and $M_\mathrm{b}$. Due to a strong greenhouse (or blanketing) effect from the steam atmosphere, most cases lead to $T_\mathrm{b} > 2000$ K. As previously stated, this consequence discards any EoS that does not hold for such high temperatures. A second observation is that at low temperatures, one input irradiation temperature $T_\mathrm{irr}$ can correspond to two different planet temperatures $T_\mathrm{p}$ (and atmospheric properties). Since our work focuses on highly irradiated exoplanets, we will only investigate cases with $T_\mathrm{irr}~>~400$ K to bypass this degeneracy.

\begin{figure}[ht!]
	\resizebox{\hsize}{!}{\includegraphics[angle=0,width=5cm]{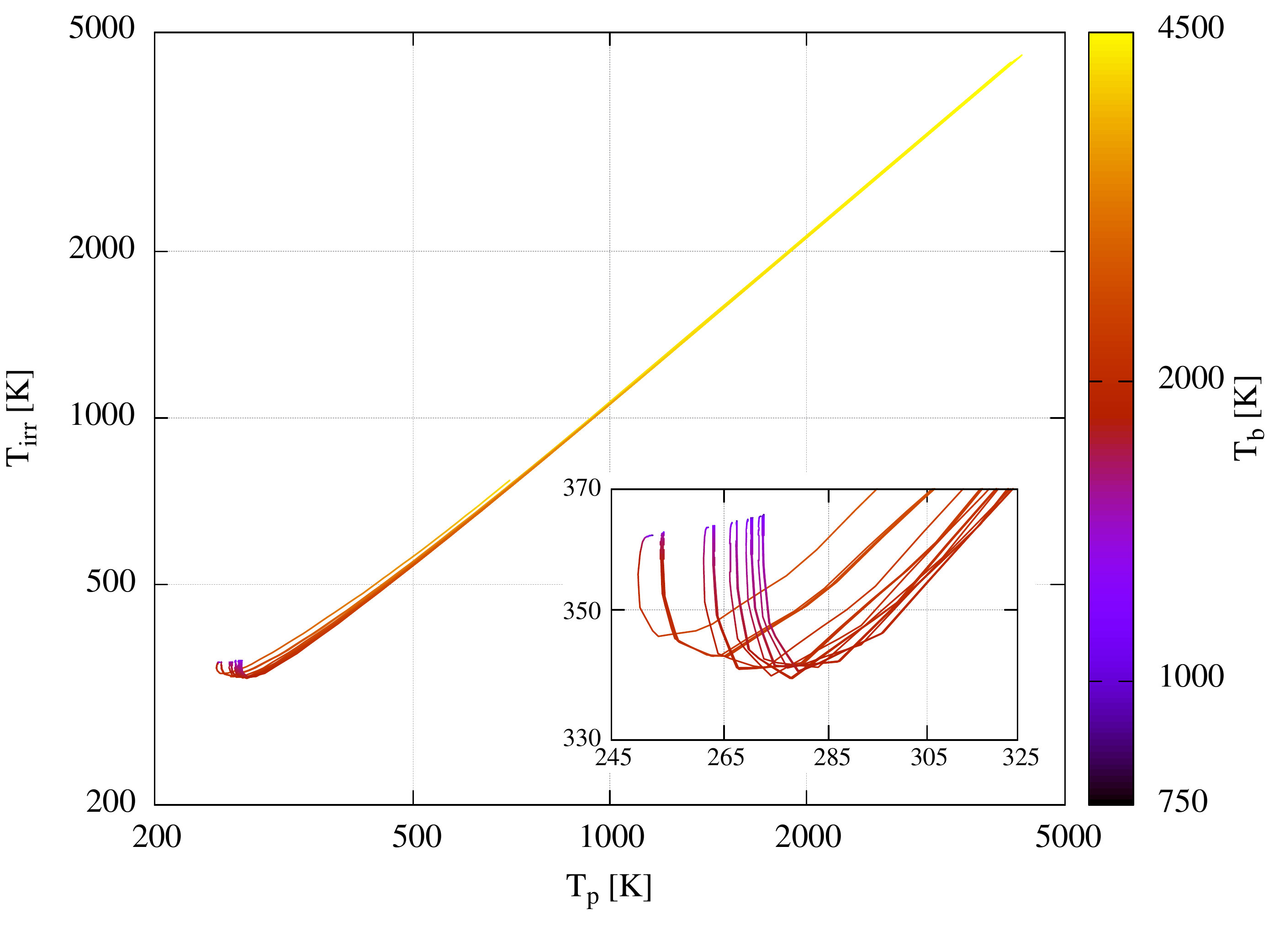}}
	\caption{Irradiation temperature $T_\mathrm{irr}$ as a function of the planet's temperature $T_\mathrm{p}$. Several curves are obtained due to different values of $g_\mathrm{b}$ and $M_\mathrm{b}$ in the available parameter range. Coldest planets exhibit a degeneracy, as the same amount of irradiation is consistent with two different atmospheric structures. As shown by the color bar, when taking the ``hot" solution for $T_\mathrm{p}$, the temperature at the bottom of the atmosphere $T_\mathrm{b}$ is $>2000$ K.}
	\label{fig:teq_tp}
\end{figure}

\section{Atmospheric escape} \label{sec:atmos-escape}

Planetary atmospheres are subject to two types of instabilities : hydrostatic and thermal escape. The former is encountered when the gravity at a given height is insufficient to retain the gas. In this case, the atmosphere cannot exist in hydrostatic equilibrium and atmospheric models fail to produce static $(P,\rho)$ profiles. The choice of $g_\mathrm{b}>3$ m.s$^{-2}$ is arbitrary, but allows to avoid these cases. The latter occurs when the thermal energy of gas molecules exceeds the gravitational potential, allowing their escape. Escape rates are then computed, indicating which molecules can remain in an atmosphere. Several mechanisms of non-thermal escape exist as well, involving collisions between atoms and ions producing kinetic energy that leads to knock-off \citep{Hu82}, but they rely on processes that are beyond the scope of this study.

\subsection{Jeans' escape}

One widely known process of atmospheric escape is the Jeans escape. Gas molecules have a velocity distribution given by the Maxwell-Boltzmann distribution, which displays an infinite extension in the velocity space, meaning that \textit{some} particles have velocities greater than the escape velocity. By integrating this distribution, one can derive the Jeans' particle flux (particles per time unit per surface unit) escaping the atmosphere at the exobase \citep{Je25}:

\begin{eqnarray}
	\Phi_J = \frac{n_\mathrm{e} v_\mathrm{esc}}{2 \sqrt{\pi}} \frac{1}{\sqrt{\lambda}} \left(1+\lambda\right) \mathrm{e}^{-\lambda}, \label{eq:jflux1}
\end{eqnarray}

\noindent where $n_\mathrm{e}$ is the particle number density at the top of the atmosphere (exobase), $v_\mathrm{esc} = \sqrt{2 g_\mathrm{b} R_\mathrm{p}}$ is the escape velocity (we assume $R_\mathrm{p}\simeq R_\mathrm{b}$ and $M_\mathrm{p}\simeq M_\mathrm{b}$). $\lambda = \left(\frac{v_\mathrm{esc}}{v_\mathrm{th}}\right)^2$ is the escape parameter, with $v_\mathrm{th}=\sqrt{2R_g T_\mathrm{e}/\mu}$ the average thermal velocity of molecules of mean molar mass $\mu$ at the exobase temperature $T_\mathrm{exo}$, and $R_g$ is the ideal gas constant. 

We wish to provide an estimate of the physical characteristics of the planets that would lose more than a fraction $x_\mathrm{lost}=0.1$ of water content over a typical timescale of $\Delta t= 1$ Gyr. This condition is met when

\begin{eqnarray}
4\pi R_\mathrm{p}^2 \frac{\mu}{\mathcal{N}_\mathrm{A}} \Phi_J \ge \frac{x_\mathrm{lost} M_\mathrm{p}}{\Delta t}, \label{eq:Jescape0}
\end{eqnarray}

\noindent with $\mathcal{N}_\mathrm{A}$ the Avogadro number. Solving Eq. (\ref{eq:Jescape0}) with Earth's properties ($n_\mathrm{e} = \frac{P_\mathrm{top} \mathcal{N}_\mathrm{A}}{Rg T_\mathrm{exo}} \sim 10^{19}$, $R_\mathrm{p}=R_\Earth$,  $M_\mathrm{p}=M_\Earth$) yields $\lambda \le 100$. Due to the exponential term, the result is poorly sensitive to changes in parameters, including the exact location of the exobase. Assuming $T_\mathrm{exo}=T_\mathrm{irr}$, this condition can be rewritten as

\begin{eqnarray}
	R_\mathrm{p} > \frac{1}{\lambda} \frac{G \mu}{R_g T_\mathrm{irr}} M_\mathrm{p}, \label{eq:Jescape}
\end{eqnarray}

\noindent with $G$ the gravitational constant. This estimate is consistent with today's composition of planets of the solar system (see Fig. \ref{fig:final}). Equation \ref{eq:Jescape} gives an indication of the properties of the planets that are subject to H$_2$ or H$_2$O escape, implying that their atmospheres should be dominated by heavier molecules (H$_2$O, CO$_2$, O$_2$, CH$_4$, etc) or be rocky planets, respectively.

\subsection{Hydrodynamic escape}

Hydrodynamic escape, also referred to as hydrodynamic blowoff, occurs when upper layers of the atmosphere are heated by intercepting the high energy irradiation (Far UV, Extreme UV and X-ray fluxes, the sum of which is often called XUV flux) from the host star. This heating induces an upward flow of gas, leading to mass-loss at a rate \citep{Er07,Ow13}

\begin{eqnarray}
\dot{M} = \epsilon \frac{L_\mathrm{XUV} R_\mathrm{p}^3}{G M_\mathrm{p} (2a)^2}, \label{eq:Hescape}
\end{eqnarray}

\noindent where $L_\mathrm{XUV}$ is the host star XUV luminosity, $a$ is the planet's orbital distance and $\epsilon$ is a conversion factor between incident irradiation energy and mechanical blowoff energy. Note that Eq. (\ref{eq:Hescape}) is only true in the energy-limited case. Heating occurs by absorption of high-energy photons by molecules which are dissociated in the upper atmosphere, meaning that blowoff can be limited by i) the number of photons as 1 photon breaks 1 molecule, and ii) recombination time as a dissociated molecule may recombine before being able to absorb the XUV irradiation again. Boundaries between these regimes have been explored by \cite{Ow16}, who showed that the sub-Neptune population undergoes mostly energy-limited mass-loss, validating the use of Eq. \ref{eq:Hescape} in our case.

For our estimate, we use the X-ray and UV luminosities obtained by fits on observational data for M to F type stars by \cite{Sa11}:

\begin{eqnarray}
&L_\mathrm{EUV} = 10^{3.8} L_\mathrm{X}^{0.86},&\\
&L_\mathrm{X} = 6.3\times 10^{-4} L_\star,& \qquad \tau<\tau_\mathrm{sat} \\
&\phantom{L_\mathrm{X}xxx} = 1.89\times 10^{21} \tau^{-1.55},& \qquad \tau>\tau_\mathrm{sat} \nonumber
\end{eqnarray}

\noindent where $\tau$ is the host star age in Gyr and $\tau_\mathrm{sat} 
= 5.72\times 10^{15} L_\star^{-0.65}$ \citep{Sa11}. To estimate the XUV luminosity, the star's bolometric luminosity is assumed constant, a hypothesis supported by the stellar evolution tracks of \cite{Ba15}. Integrating the XUV luminosity in the saturation regime ($0<\tau < \tau_\mathrm{sat}$) and beyond, gives the finite quantity $E_\mathrm{XUV}=\int_{0}^{+\infty} L_\mathrm{XUV}~dt = 1.8 \times 10^{39} $ W for a solar type star.

Again, we look for planets that could lose more than 10\% of their mass over a 1 Gyr period, due to atmospheric blowoff:

\begin{eqnarray}
\epsilon \frac{E_\mathrm{XUV} R_\mathrm{p}^3}{G M_\mathrm{p} (2a)^2} \ge x_\mathrm{lost} M_\mathrm{p}. \label{eq:Hescape1}
\end{eqnarray}

\noindent Combining Eq. (\ref{eq:tirr_obs}) and Stefan-Boltzmann's law $L_\star~=~4\pi R_\star^2 \sigma_\mathrm{sb} T_\mathrm{eff}^4$ gives
\begin{eqnarray}
	(2a)^2 = \frac{1}{T_\mathrm{irr}^4} \frac{L_\star}{4 \pi \sigma_\mathrm{sb}}.
\end{eqnarray}
Substituting this expression in Eq. (\ref{eq:Hescape1}) yields to the condition:

\begin{eqnarray}
R_\mathrm{p} \ge M_\mathrm{p}^{\frac{2}{3}} \left(\frac{x_\mathrm{lost} G}{\epsilon 4\pi \sigma_\mathrm{sb} T_\mathrm{irr}^4 E_\mathrm{XUV}}\right)^{\frac{1}{3}}. \label{eq:Hescape2}
\end{eqnarray}

\noindent This condition only gives an indication of the planets that are subject to substantial hydrodynamic escape. All arbitrary quantities such as $\epsilon\simeq1$ \citep{Ow12,Bo17} and $E_\mathrm{XUV}$, are affected by a power of $1/3$, resulting in a low dependency on the chosen values.

The nature of escaping particles is not considered in Eq. (\ref{eq:Hescape2}), meaning the computed quantity is the total lost mass. \cite{Bo17} developed a method to quantify the hydrodynamic outflow $r_\mathrm{F}$ (how many atoms of oxygen leave for each hydrogen atom). Based on their work, we compute $r_\mathrm{F}\sim 0.2$, indicating substantial loss of both H and O, with an accumulation of O$_2$. Mass loss of water content and accumulation of O$_2$ have several implications for the habitability of exoplanets \citep{Ri16,Sc16}. The power laws for mass is 1 and $\frac{2}{3}$ in the cases of Eqs. (\ref{eq:Jescape}) and (\ref{eq:Hescape2}), respectively. This implies that hydrodynamic escape is more efficient for less dense planets. In contrast, Jeans escape is dominant in the case of denser planets. The power-law for $T_\mathrm{irr}$ is $-1$ and $-\frac{4}{3}$ in the cases of Eqs. \ref{eq:Jescape} and \ref{eq:Hescape2}, respectively, implying that hydrodynamic escape will take over Jeans escape at higher irradiation temperatures. As shown in Fig. \ref{fig:mr+}, Eqs. \ref{eq:Jescape} and \ref{eq:Hescape2} leave a window for planets that lost their H$_2$ reservoir but kept heavier volatiles from which they formed \citep{Ze19}. This result highlights the consistency between the possible existence of irradiated ocean planets and atmospheric escape.

\section{Results} \label{sec:results}

\begin{figure*}[!ht]
	\resizebox{\hsize}{!}{\includegraphics[angle=0,width=5cm]{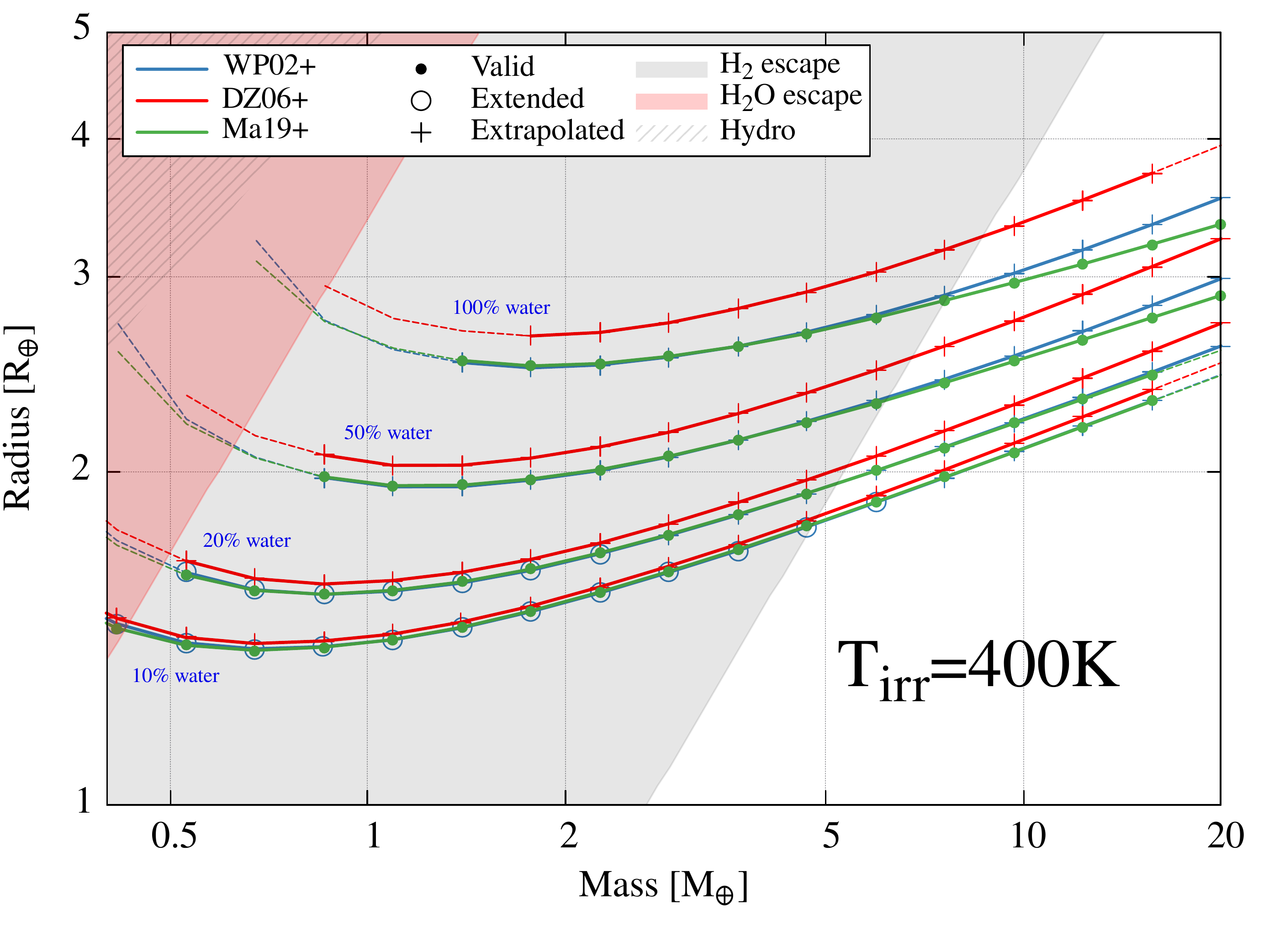} \includegraphics[angle=0,width=5cm]{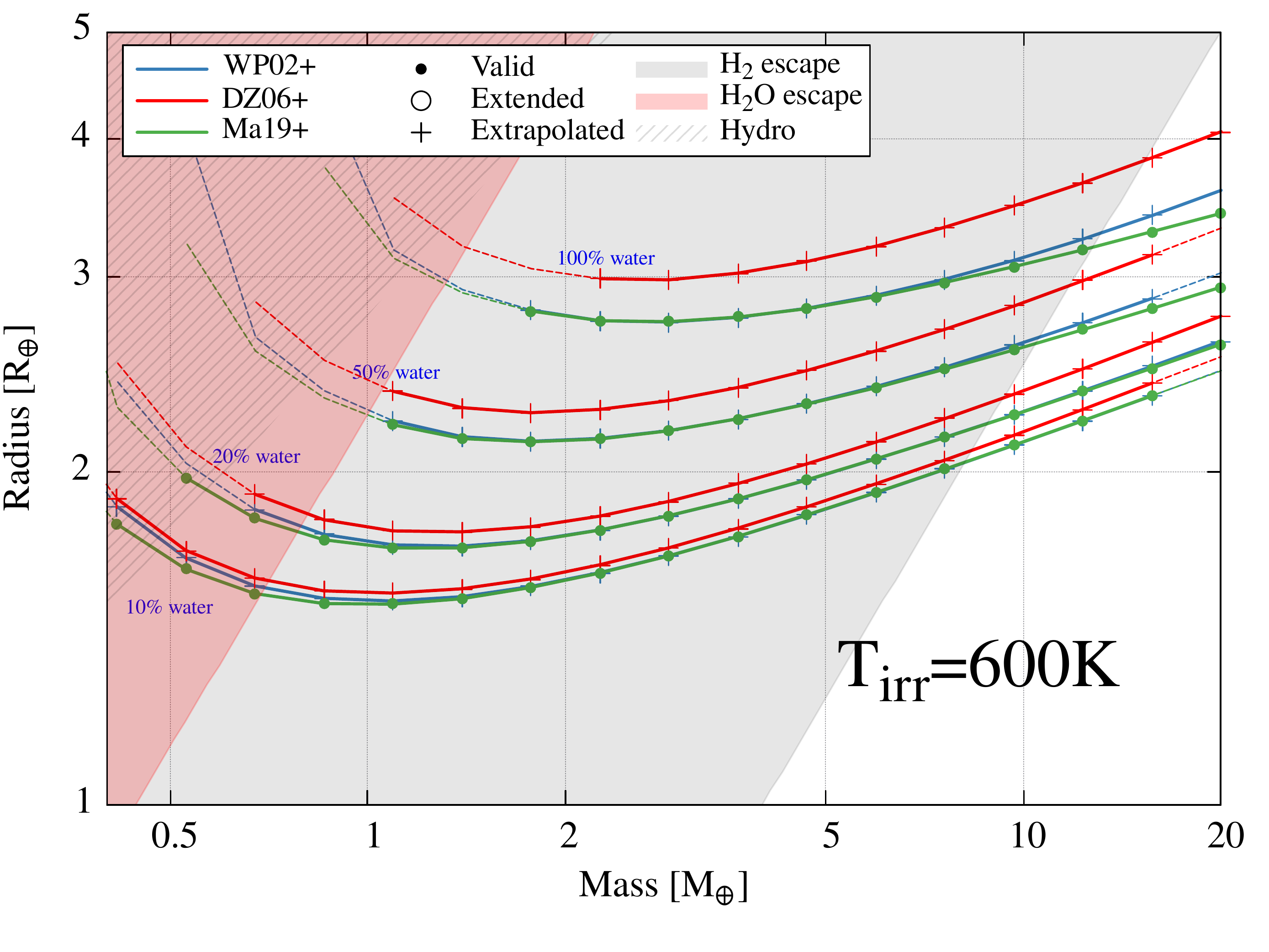}}
	\resizebox{\hsize}{!}{\includegraphics[angle=0,width=5cm,]{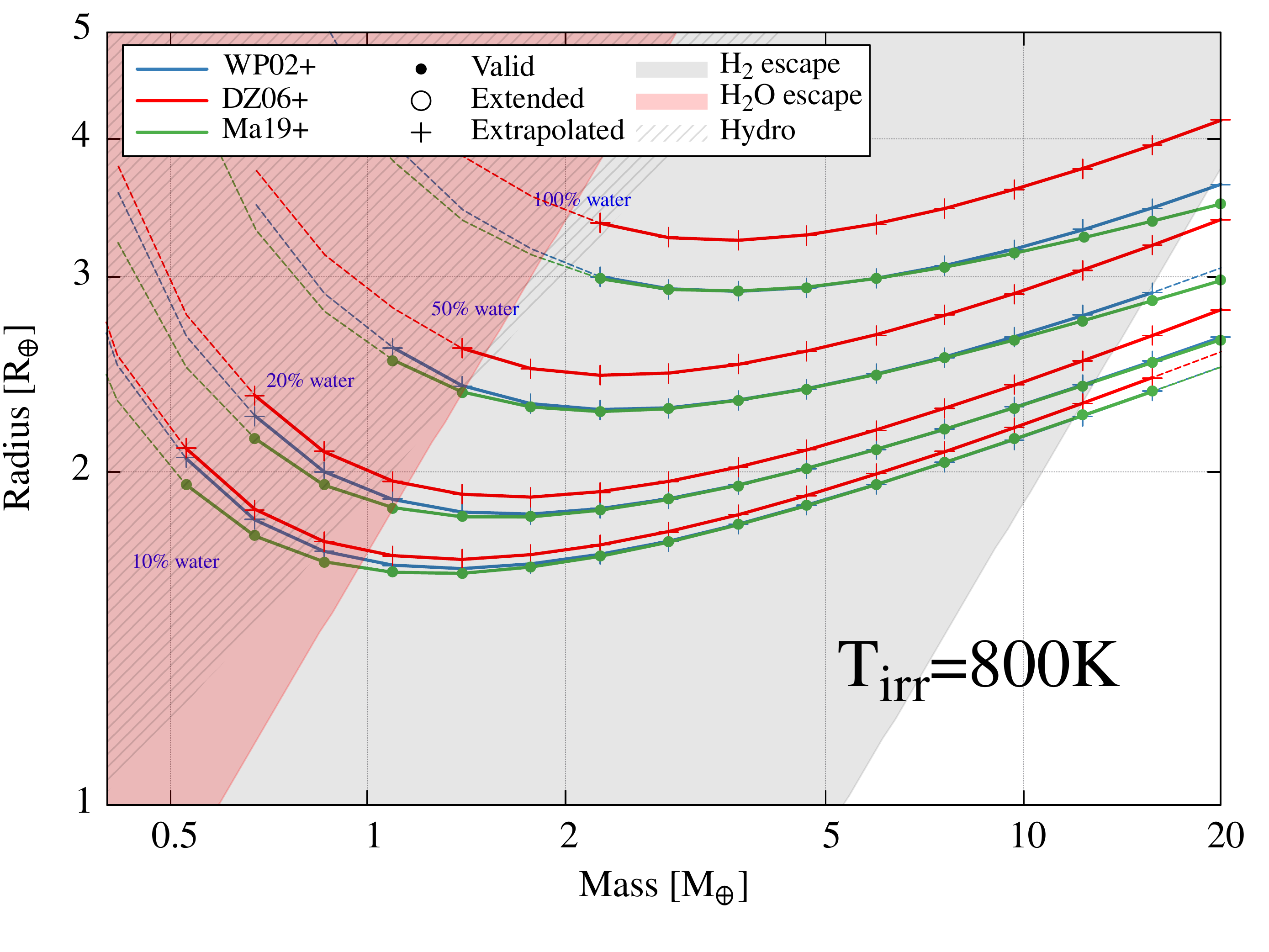} \includegraphics[angle=0,width=5cm]{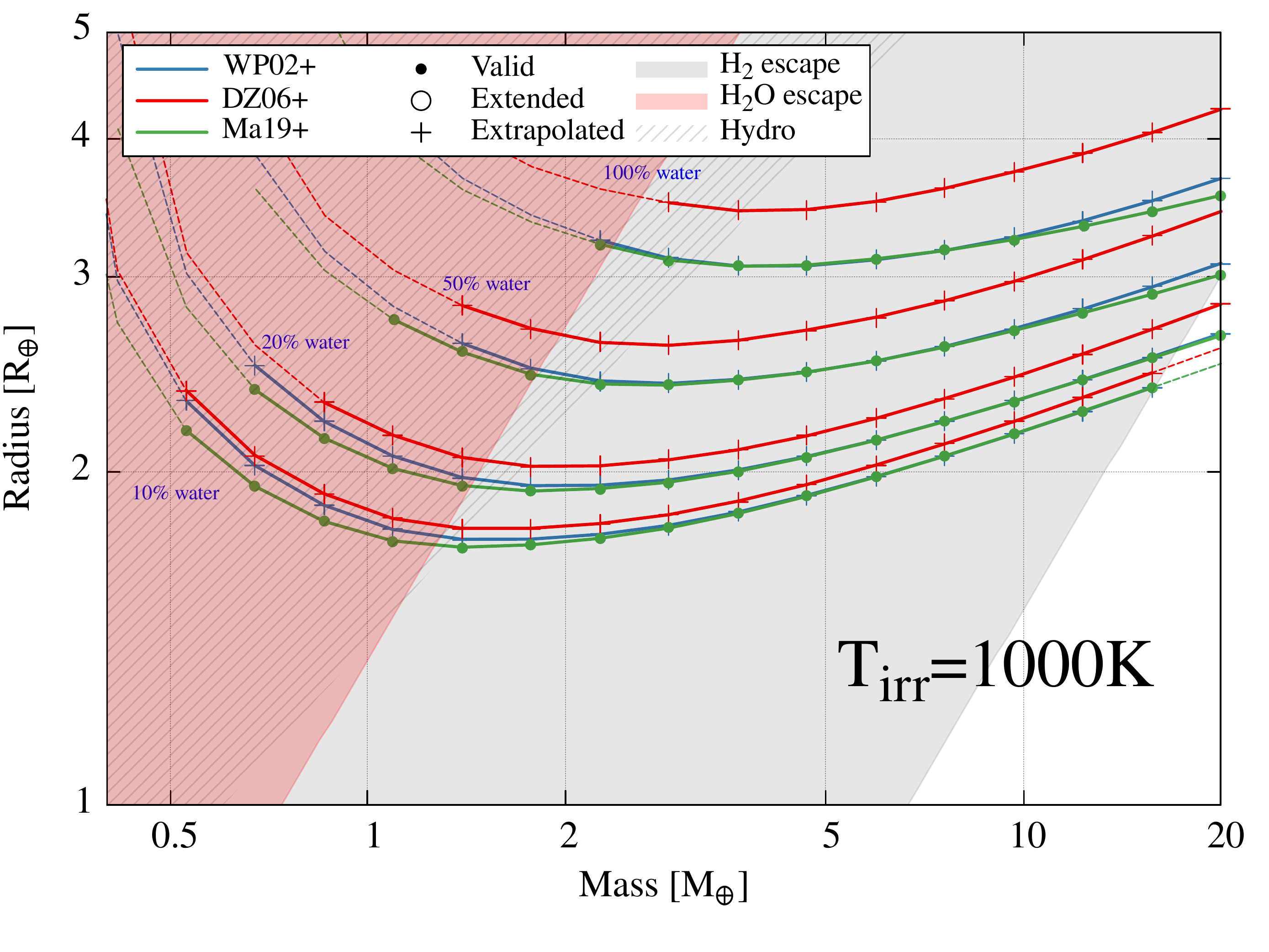}}
	\caption{Mass-radius relationships for planets with Earth-like properties regarding their rocky part (see Table \ref{tab:parameters}), and computed with $\gamma_+$, for multiple temperatures and water contents. Colors correspond to the three used EoSs: DZ06 (red), WP02 (blue) and Ma19 (green). Dashed lines correspond to regions where the atmosphere model is extrapolated beyond the available grid (see Appendix \ref{sec:trilinear}). Filled circles correspond to cases where both $P$ and $\gamma$ remain in the range of validity of used EoS. Open circles correspond to cases where $P$ or $\gamma$ are computed in the extended range. Crosses correspond to cases where $P$ or $\gamma$ are in the extrapolated range. Shaded areas correspond to H$_2$ (gray), H$_2$O (pink) and hydrodynamic escape (shaded) (see Sec. \ref{sec:atmos-escape}).}
	\label{fig:mr+}
\end{figure*}

\begin{figure*}[!ht]
	\resizebox{\hsize}{!}{\includegraphics[angle=0,width=5cm]{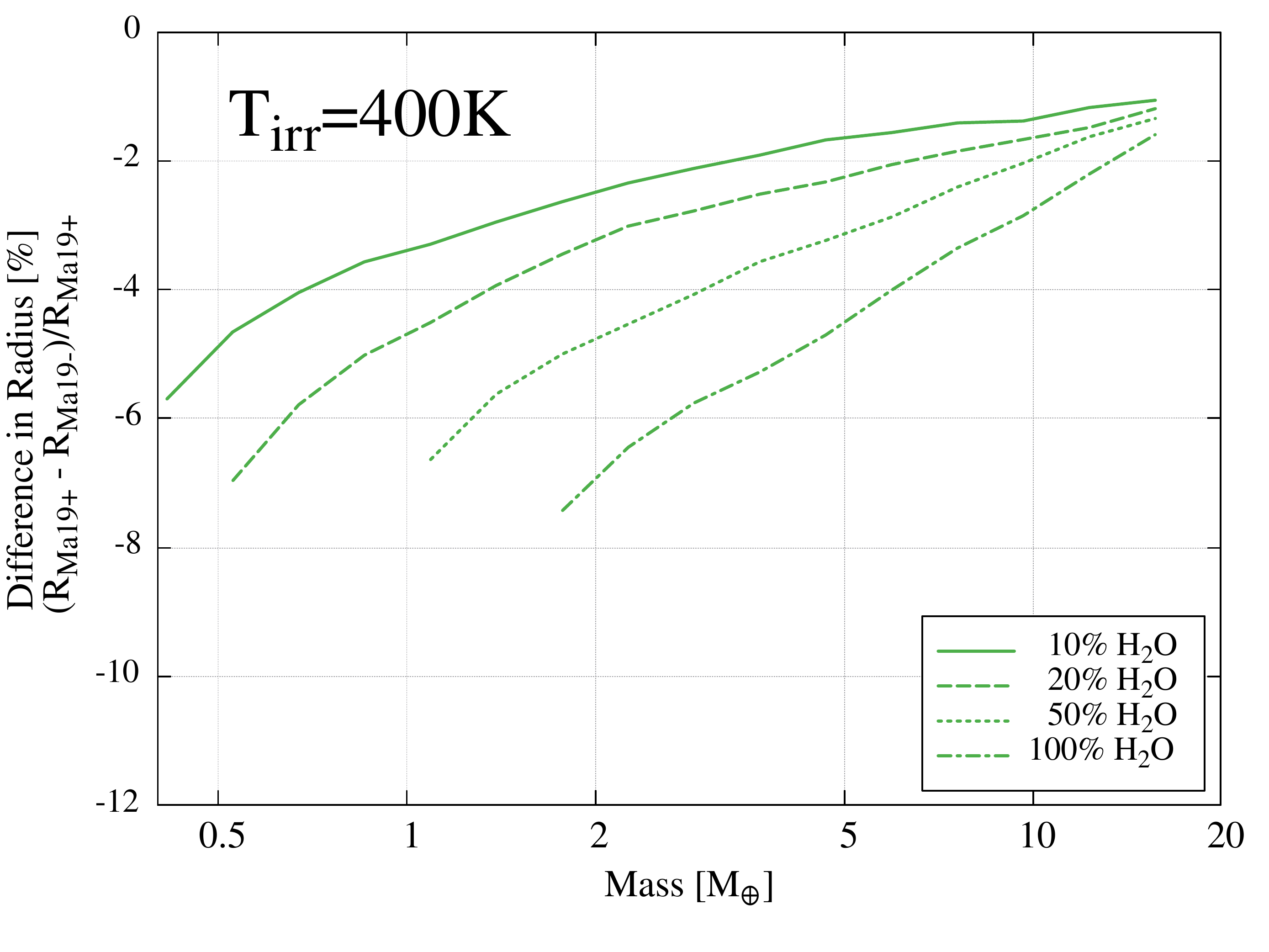} \includegraphics[angle=0,width=5cm]{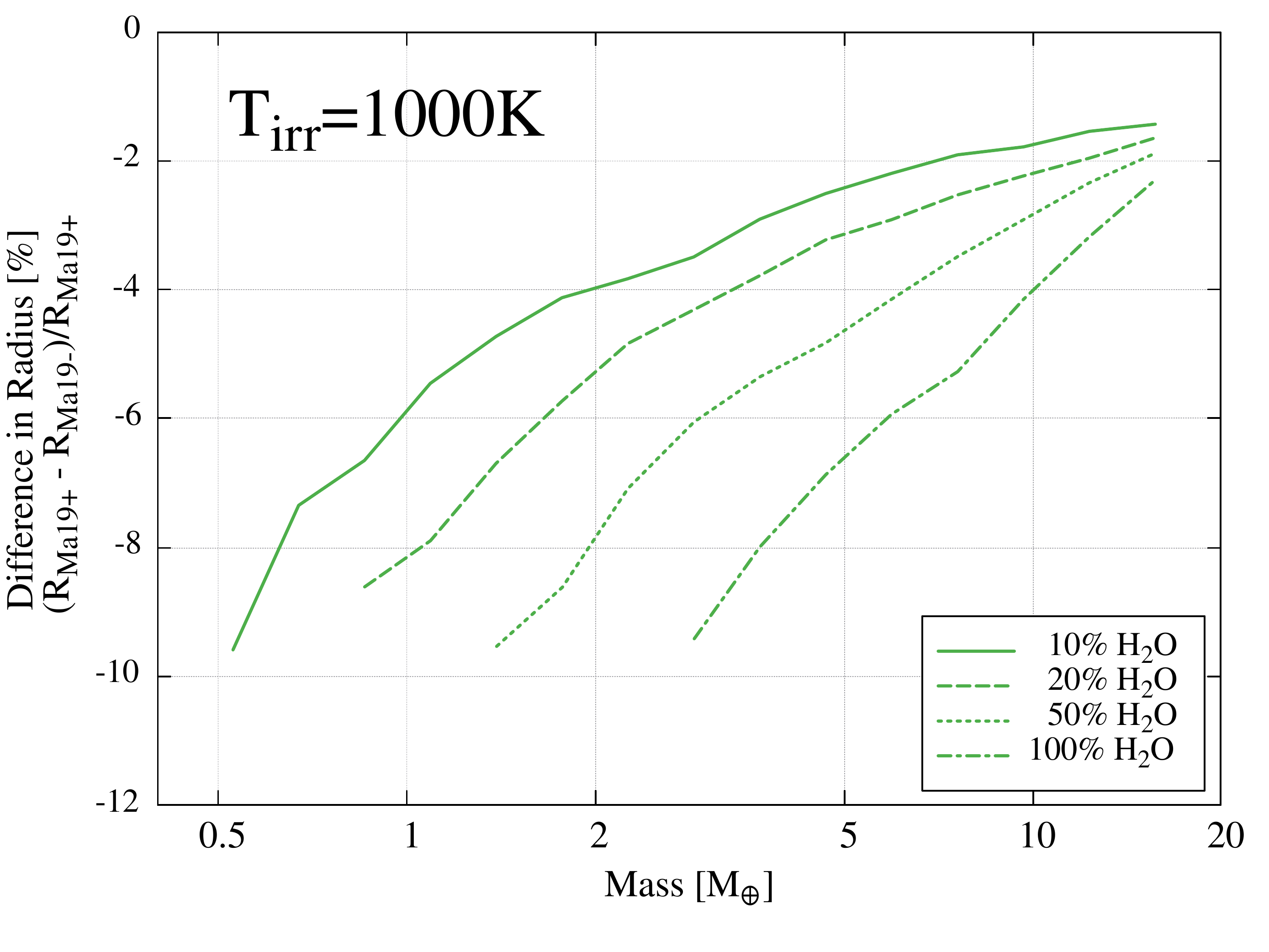}}
	\caption{Relative difference on radius between Ma19+ and Ma19- parametrizations showing the large impact of the temperature profile on mass-radius relationships.}
	\label{fig:mdr}
\end{figure*}

The aim of this paper is to quantify the impact of the choice of EoS and $\gamma$ computation on mass-radius relationships. With three EoSs and two $\gamma$ parametrizations, 6 cases are considered : WP02$\pm$, DZ06$\pm$ and Ma19$\pm$, with the +/- sign standing for $\gamma_+$ and $\gamma_-$, respectively. For each case, three validity domains are explored: true validity range, extended range, or extrapolated range. A case is valid when the $(P,T)$ profile remains strictly in the true validity range of the used EoS and $\gamma$ computation. It is extended if the EoS and/or $\gamma$ computation reaches the extended range. If either the EoS or $\gamma$ reaches the extrapolated region, the whole case is considered extrapolated. For example, the Ma19+ parametrization (see Fig. \ref{fig:mr+}) is always valid due to the important validity range of Ma19 EoS and $\gamma_+$ computation. On the other hand, the Ma19- parametrization is always extrapolated because the computed $\gamma_-$ is out of its validity and extended range.

\subsection{Mass-radius relationships and choice of EoS}

Figure \ref{fig:mr+} presents computed mass-radius relationships for the $\gamma_+$ parametrization, and assuming Earth-like properties for the rocky part (see Table \ref{tab:parameters}). As predicted from the shape of EoSs curves, WP02 and DZ06 EoSs underestimate the density and thus produce larger planets. This effect is accentuated for more massive planets with a larger amount of water, corresponding to cases where water pressure reaches the highest values. The radius is also overestimated for low-mass planets, because the hydrosphere becomes extended due to the low gravity, implying that a slight underestimation of the density can still lead to a substantial difference in radius. These results show the incontestable asset of the EoS developed by \cite{Ma19}, and rule out the possibility of using WP02 or DZ06 EoSs to produce reliable mass-radius relationships for planets with substantial amounts of water. To remain in the true validity ranges of WP02 or DZ06 EoS, one should consider a few \% of water content at most in the planet.

As discussed in Sec. \ref{sec:gruneisen}, $\gamma_+$ is always lower than $\gamma_-$ in Earth-sized planets fully made of water. As a result $(P,T)$ profiles for $\gamma_+$ parametrizations are steeper than for $\gamma_-$ parametrizations (see Fig. \ref{fig:gruneisen_val}), meaning the interior is colder for $\gamma_+$. In turn, colder planets will be denser and thus smaller. The impact of the choice between $\gamma_+$ and $\gamma_-$ is shown in Fig. \ref{fig:mdr}, where the relative difference on the radius between Ma19+ and Ma19- parametrizations is presented. In all cases, the relative difference between the models is 10\%  at most.

As the mass of a planet increases, its gravity becomes more important, and its hydrosphere (interior structure and atmosphere) consequently thinner. Thinner hydrospheres, especially in the case of massive planets, lead to smaller relative differences in radii. Moreover, values of $\gamma_+ $ and $ \gamma_-$ become closer (and even equal) in the 10$^1$--10$^2$ GPa pressure range (see Fig. \ref{fig:gruneisen_val}), thus reducing even more significantly the radii differences between the Ma19+ and Ma19- parametrizations.

The value of $\gamma$ increases when the ($P$,$T$) curves of a hydrosphere approaches the liquid--Ice VII transition, which leads to a more important temperature gradient that prevents the formation of high pressure ices. This observation is in major disagreement with models assuming isothermal hydrospheres \citep{Va06,Va07,Se07,Ze13,Br17,Ze19}, an hypothesis often justified by assuming that temperature has a secondary impact on EoSs, 
which remains a valid statement  for solid phases but not in the case of the hydrosphere. A correct treatment of the temperature gradient \citep{Mo20} leads to the presence of high-temperature phases for H$_2$O (ionic, super ionic, plasma), which are more dilated, impacting significantly the mass-radius relationships.

In the following, we use $\gamma_+$ and Ma19 to compute the mass-radius relationships. Indeed, the pressure and temperature ranges in the hydrospheres of sub-Neptunes-like planets lie well in the region for which the Ma19 formulation was developed. Also, due to the blanketing effect of the atmosphere, even the coldest planets irradiated at $T_\mathrm{irr}~=~400$ K have a temperature of more than 2000 K at the 300 bar interface (see Fig. \ref{fig:teq_tp}), which corresponds to the pressure at which the atmospheric and the internal model are connected. This interface is already located well above the range of validity of $\gamma_-$.

\subsection{Planetary composition}

\begin{figure}[!ht]
	\resizebox{0.92\hsize}{!}{\includegraphics[angle=0,width=5cm]{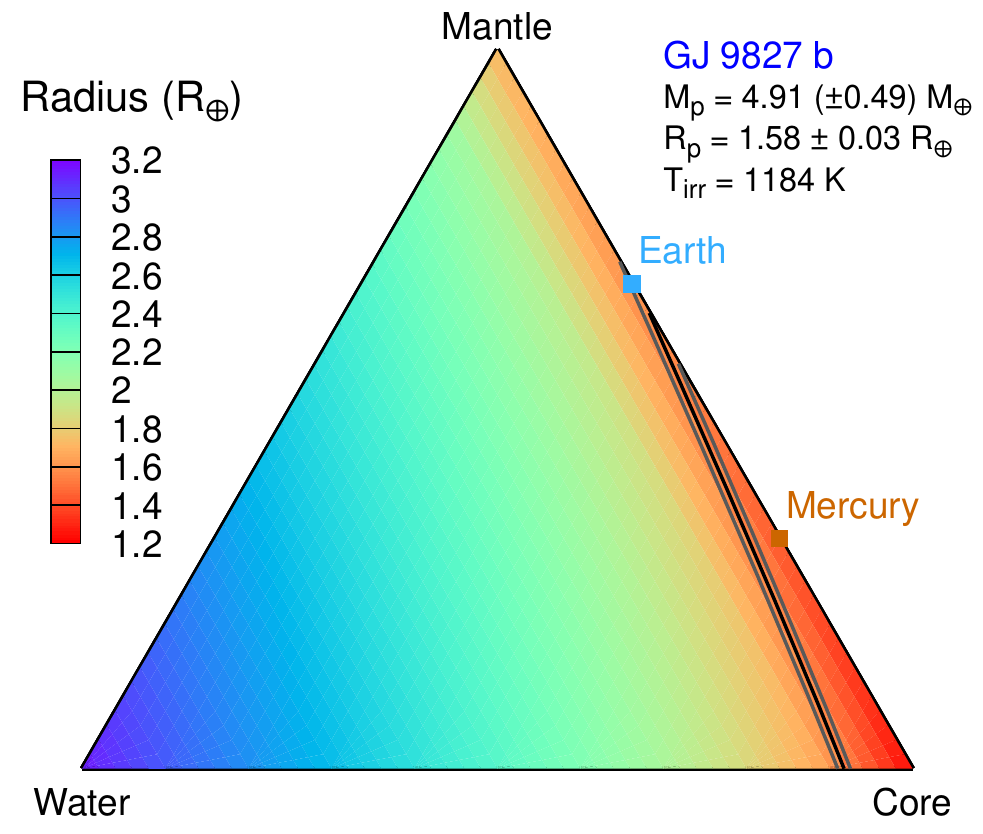}}\\
	\resizebox{0.92\hsize}{!}{\includegraphics[angle=0,width=5cm]{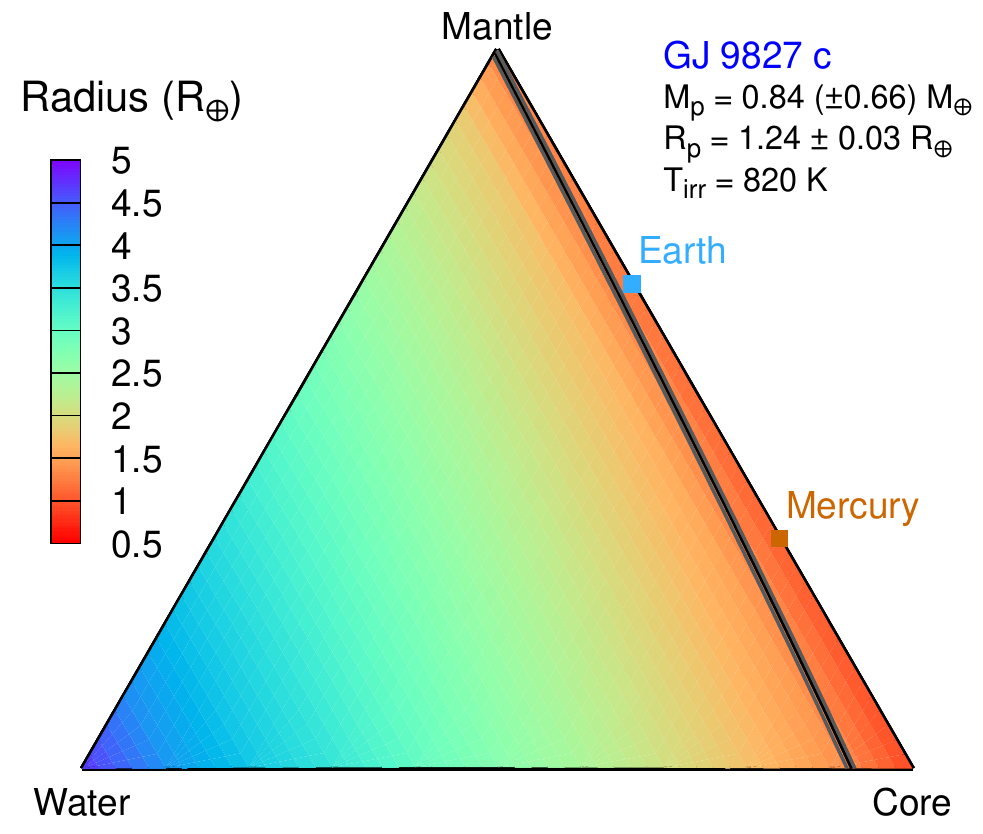}}\\
	\resizebox{0.92\hsize}{!}{\includegraphics[angle=0,width=5cm]{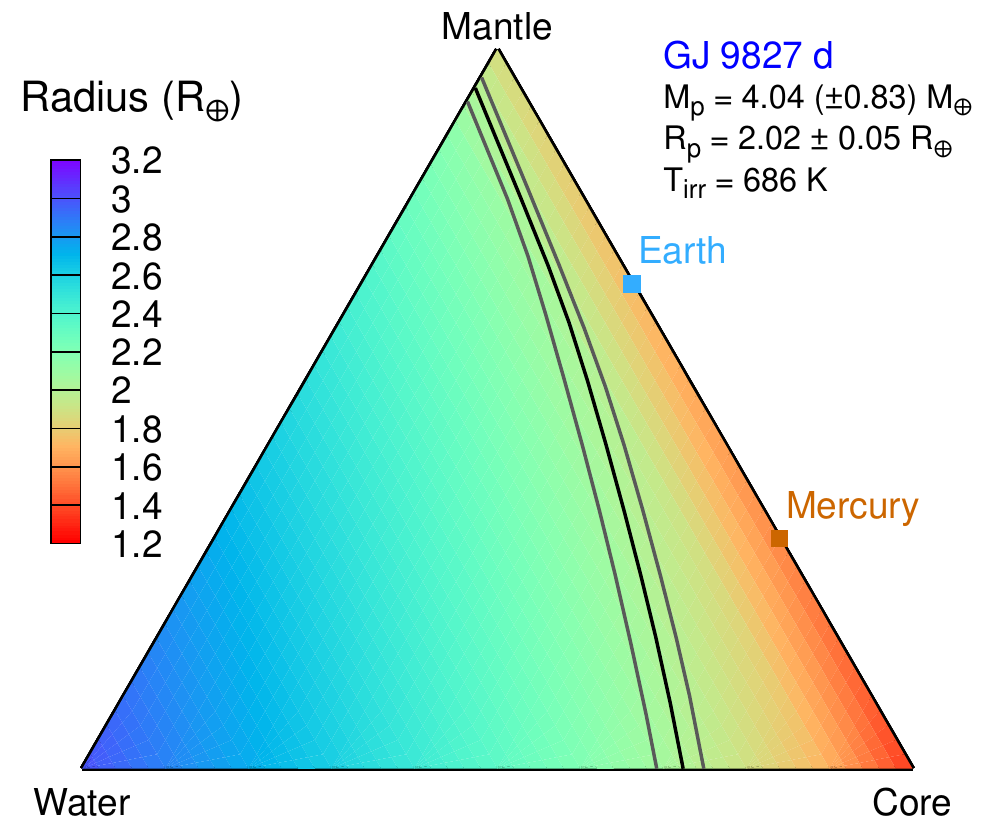}}
	\caption{From top to bottom: possible compositions of planets b, c and d of the GJ 9827 system \citep{Ri19} in the forms of compositional ternary diagrams. Ternary diagrams were computed for the central masses of the planets, and contours are plotted for the measured radius and 1$\sigma$ error bar.}
	\label{fig:ternary}
\end{figure}

Mass-radius relationships only provide an order of estimate of the possible exoplanet composition. A more precise assessment is achieved via the use of compositional ternary diagrams. For a given planet mass and irradiation temperature, such a diagram shows the radius as a function of the planet's WMF and CMF. Possible compositions as thus retrieved from the contour at the level of the planet's measured radius. Computations presented here use only the central value of the mass of each planet, thus not taking into account the measurement error on the planet's mass.

\begin{figure*}[!ht]
	\resizebox{\hsize}{!}{\includegraphics[angle=0,width=10cm]{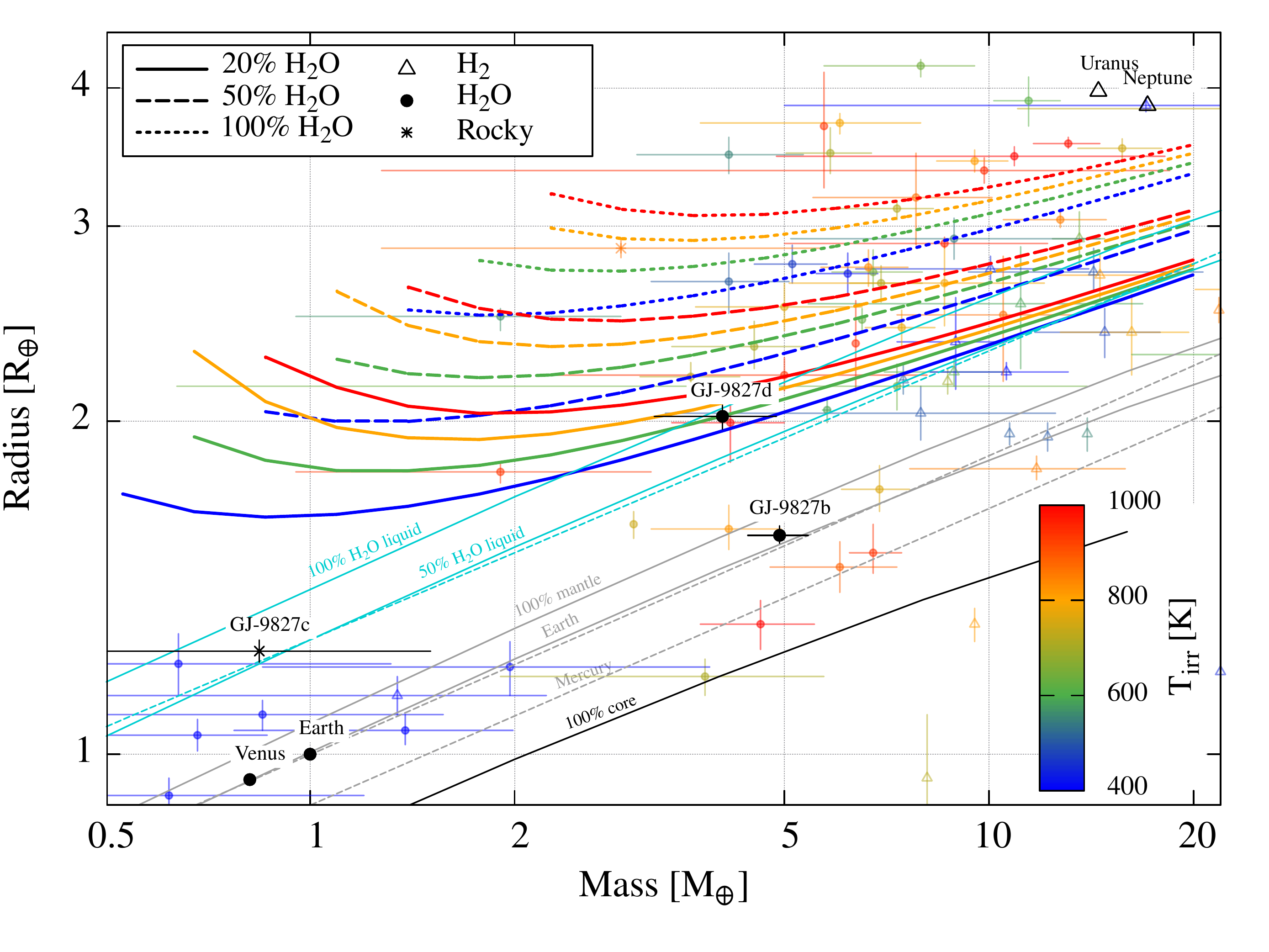}}
	\caption{Comparison between mass-radius relationships computed with the Ma19+ model and those existing in literature. Our mass-radius relationships were computed for WMF of 20\%, 50\% and 100\% with no metallic core, and temperatures of 400, 600, 800 and 1000 K. Thin solid lines and thin dashed lines are from  \cite{Ze16} and \cite{Br17}, respectively. Empty triangles, solid circles and stars correspond to planets subject to no atmospheric escape, to escape of H$_2$ only and to escape of both H$_2$ and H$_2$O (Jeans or blowoff), respectively. Planetary data are taken from the NASA exoplanet archive and updated to July 2020.}
	\label{fig:final}
\end{figure*}

Possible compositions of the three planets of the GJ 9827 system are shown in Fig. \ref{fig:ternary}, based on the planets parameters measurements made by \cite{Ri19}. Planet b exhibits an Earth-like interior without the need of invoking a significant steam atmosphere. The presence of a thick steam atmosphere is rather consistent with the low-density measurements made for planet c, with a water content ranging from 1 to 8$\%$. Physical properties (mass, radius and temperature) of planet c lead to important Jeans' escape (with our criterion in Eq. \ref{eq:Jescape0}, see Fig. \ref{fig:final}), suggesting the absence of H$_2$ and He in the atmosphere. Moreover, planet c is unlikely to accrete substantial amount of H$_2$ and He due to its low mass. Although planet d is consistent with a Jupiter-like interior due to their similar bulk densities, again, its high irradiation temperature suggests the presence of a H$_2$-He free atmosphere. Isochrones used by \cite{Ri19} fix a lower limit on the age of 5 Gyr on the age of the system, which makes an H$_2$-He atmosphere less likely as Jeans' escape would remove them. Applying our model to the current measurements yields a WMF in the 5--30\% range for planet d. These results are summarized in Table \ref{tab:gj}.

\begin{table}
	\centering
	\caption{Planetary parameters of the GJ 9827 system used as input for the model, and estimated WMF using ternary diagrams (Fig. \ref{fig:ternary}).} 
	\label{tab:gj}
	\begin{tabular}{llll}
		\tablewidth{0pt}
		\hline
		\hline
		Planet 	& b 		& c & d 	\\	
		\hline
		$M_\mathrm{p} ~(M_\Earth)$ & $4.91\pm0.49$ & $0.84\pm0.66$ & $4.04\pm0.83$ \\ 	
		$R_\mathrm{p}~(R_\Earth)$ & $1.58\pm0.03$ & $1.24\pm0.03$ & $2.02\pm0.05$ \\
		$T_\mathrm{irr}$ (K) & 1184 K & 820 K & 686 K \\ \hline
		WMF (\%) & 0--5 & 1--5 & 5--30 \\
		\hline
	\end{tabular}
\end{table}

Ternary diagrams presented here do not take into account the uncertainty on each planet's mass, and were computed for the central value only. If a planet's mass is slightly higher (resp. lower), its density increases (resp. decreases), while the estimated WMF diminishes (resp. grows). This implies that the mass and radius of a planet must be measured with extreme accuracy to constrain the WMF properly. Additional constraints can be applied from observational data such as the stellar elemental ratios (Fe/Si, Mg/Si) that could help constraining the core to mantle mass ratio \citep{Br17}, and methods such as MCMC can be performed to simultaneously determine all parameters \citep{Ac21}.

Figure \ref{fig:final} represents the computed mass-radius relationships for WMF of 0.2, 0.5 and 1. In this figure, the condition for substantial atmospheric loss due to Jeans' escape is derived by solving equation (\ref{eq:Jescape0}) for each planet. One already known effect is that steam atmospheres are very extended \citep{Mo20}, allowing to compute compositions without invoking small H$_2$-He enveloppes (1-5\% by mass). The second effect is heating due to the adiabatic gradient, which decreases the density, and then increases the radius. In the 10--20 M$_\Earth$ range, the radius of a planet with a WMF of 50\% made of liquid H$_2$O is equal to that of a planet with a WMF of 20\% constitued of supercritical H$_2$O. Also, the radius of a planet fully made of liquid H$_2$O is equivalent to that of a planet with half of its mass constituted of supercritical H$_2$O. This shows how important the error on the computation of WMF can be, depending on the physical assumptions made. In the figures presented in \cite{Mo20}, where the DZ06 EoS was used, the model was able to match Neptune's mass (17 $M_\Earth$) and radius (3.88 $R_\Earth$) with a $95\%$ H$_2$O interior at 300 K. With the Ma19 EoS, a 100\% water planet presents a radius of 3.25 $R_\Earth$ at $T_\mathrm{irr}=400$ K and 3.6 $R_\Earth$ at $T_\mathrm{irr}=1300$ K.

\subsection{Analytical expression of mass-radius relationships}

All produced mass-radius relationships are very well approximated by an equation of the form
\begin{eqnarray}
	\log R_\mathrm{p} = a \log M_\mathrm{p}+b
	+\exp \left(-d(\log M_\mathrm{p}+c)\right), \label{eq:fit1}
\end{eqnarray}

\noindent where log denotes the decimal logarithm, and $R_\mathrm{p}$ and $M_\mathrm{p}$ are normalized to Earth units. $a$, $b$, $c$ and $d$ are coefficients obtained by fits, and have one value for each composition $(x_\mathrm{core},x_{\mathrm{H}_2 \mathrm{O}})$ and each temperature $T_\mathrm{irr}$. For each fitted curve, we define the mean absolute error between data and fit as
\begin{eqnarray}
	\mathrm{MAE} = \frac{1}{N}\sum_{i=1}^{N} \left|\frac{R_\mathrm{p,model}-R_\mathrm{p,fit}}{R_\mathrm{p,model}}\right|,
\end{eqnarray}
Values of the MAE are 0.01--1\% for all fits, indicating a good accuracy. The largest deviation between one point $(M_\mathrm{p},R_\mathrm{p})$ and the fitted curve is of $2.3\%$, meaning the deviation between data and fit can be neglected. Fitted coefficients vary smoothly with respect to the three parameters $(x_\mathrm{core},x_{\mathrm{H}_2 \mathrm{O}},T_\mathrm{irr})$, allowing a good interpolation of the intermediate values. The produced grid uses the compositional parameters for the core and mantle calibrated for Earth (see Table \ref{tab:parameters}), and data may be different if Fe/Si or Mg/si ratios are different.

\section{Discussion and conclusion} 
\label{sec:ccls}

This work aimed at describing a model that computes a realistic structure for water-rich planets. This was achieved by combining an interior model with an updated EoS for water, and an atmospheric model that takes into account radiative transfer.

Various EoSs were investigated, and we find that results are identical when all of them are used within their validity range. However, the pressure profile rises sharply for planets with substantial amounts of water, invalidating the use of WP02 and DZ06 EoSs for WMF $>5\%$. The blanketing effect due to the presence of the atmosphere leads to boundary temperatures greater than $2000$ K, leaving even less room for the DZ06 EoS to work properly. Both non-valid EoSs lead to the common result of overestimating the planetary radius by up to $\sim$10\%. Inexact computation of the Gr\"uneisen parameter yields another $\sim$10\% of error on the radius, at most. This requires to use an EoS that holds for pressures up to a few TPa and temperatures of $10^4$ K (conditions at the center of a pure water sphere of 1 Jupiter mass), such as \cite{Ma19}.

Other EoSs exist in the literature, such as those proposed by \cite{Br18} and \cite{Ha20}, which are functions either fitted or derived from the Gibbs or Helmholtz free energy. The range of validity for the EoS of \cite{Br18} is less extended than that of \cite{Ma19}, justifying our choice of EoS. \cite{Ha20} presents a unified EoS for water from the connexion of already existing EoSs in their validity range, incuding \cite{Ma19}. This EOS is then consistent with ours in the range of temperature and pressure explored here. The implementation of such an EoS is interesting for future works, especially when combining high pressure ices. 

It should be noted that the most accurate EoS possible is not sufficient to produce precise mass-radius relationships for such planets. Assuming an adiabatic profile for the atmosphere (i.e. not taking into account radiative transfer) results in more extended atmospheres, as heat is transported solely by convection. Isothermal water layers seem closer to reality, but they produce the same mass-radius relationships as for liquid water \citep{Ze16,Br17,Ha20}. Atmospheric models are essential for computing the atmosphere thickness and the energy that is transported to the interior.

Derived MR relationships produce radii that match well those of the population of sub-Neptunes (1.75--3.5 $R_\Earth$). This population corresponds to the second peak of the bimodal distribution of planet radii highlighted by \cite{Fu17}, thus suggesting that irradiated ocean planets are good candidates to represent such planets \citep{Mo20}. This bimodal distribution in planet radii has been predicted by \cite{Ow13} and \cite{Lo13} who investigated the atmospheric mass loss for Jupiter-like planets. However, the authors focused mainly on the loss of the enveloppe of a H/He rich atmosphere. More recently, \cite{Ow19} pointed out the need to extend this work to steam atmospheres. Our calculations aimed to do so in a very simplistic manner. Due to its greater density, we find that water is much less subject to atmospheric escape than H/He. This suggests that highly irradiated planets could have lost their H/He content through atmospheric loss processes, and the remaining matter led to either super-Earths ($R_\mathrm{p} =$ 1--1.75 $R_\Earth$) or a sub-Neptunes ($R_\mathrm{p}=$ 1.75--3.5 $R_\Earth$), depending on the final WMF.

The data grid can be used to assess a planet’s composition once its mass and radius are known. Interpolating between the values can provide better precision. For a very precise computation, the full model is required since compositional parameters such as Fe/Si and Mg/Si ratios are required as well and depend on the star spectral analysis.

Tabulated mass-radius relationships and the coefficients obtained by fit for analytical curves can be found at \url{https://doi.org/10.5281/zenodo.4552188} or \url{https://archive.lam.fr/GSP/MSEI/IOPmodel}. Explored parameter ranges are large enough to constrain planetary compositions for any WMF and CMF, and interpolate between given values without the need for the full model. We used the GJ 9827 system as a test case for our new relationships. Measured masses and radii of planets b and c of the GJ 9827 system indicate Earth-like or Venus-like interiors. We find that planet d could be an irradiated ocean planet with a WMF of $20\pm10\%$.

In the present model, only H$_2$O as a volatile is considered. Other volatiles such as CO$_2$, CH$_4$ or N$_2$ are expected to have similar densities as H$_2$O, thus producing similar mass-radius relationships. However, using a different gas will highly impact radiative transfer. Efficient radiative transfer for gases such as N$_2$ could keep the interior cold enough for maintaining a liquid water ocean, as it is the case for the Earth. An atmosphere dominated by gases such as H$_2$O or CO$_2$ lead to important blanketing, resulting in a Venus-like case.

Atmospheric escape has motivated our focus on H/He-free atmospheres. The addition of H$_2$ to the atmosphere is the scope of future work. The addition of O$_2$ as the product of water photodissociation will be considered as well.

\section*{Aknowledgements}
OM and MD acknowledge support from CNES. We thank the anonymous referee for useful comments that helped improving the clarity of our paper and added important discussion.

\newpage
\appendix

\section{Connection of internal and atmospheric models} \label{sec:connection}

\begin{figure}[ht!]
	\includegraphics[trim=6cm 6.5cm 6.5cm 1.5cm,clip]{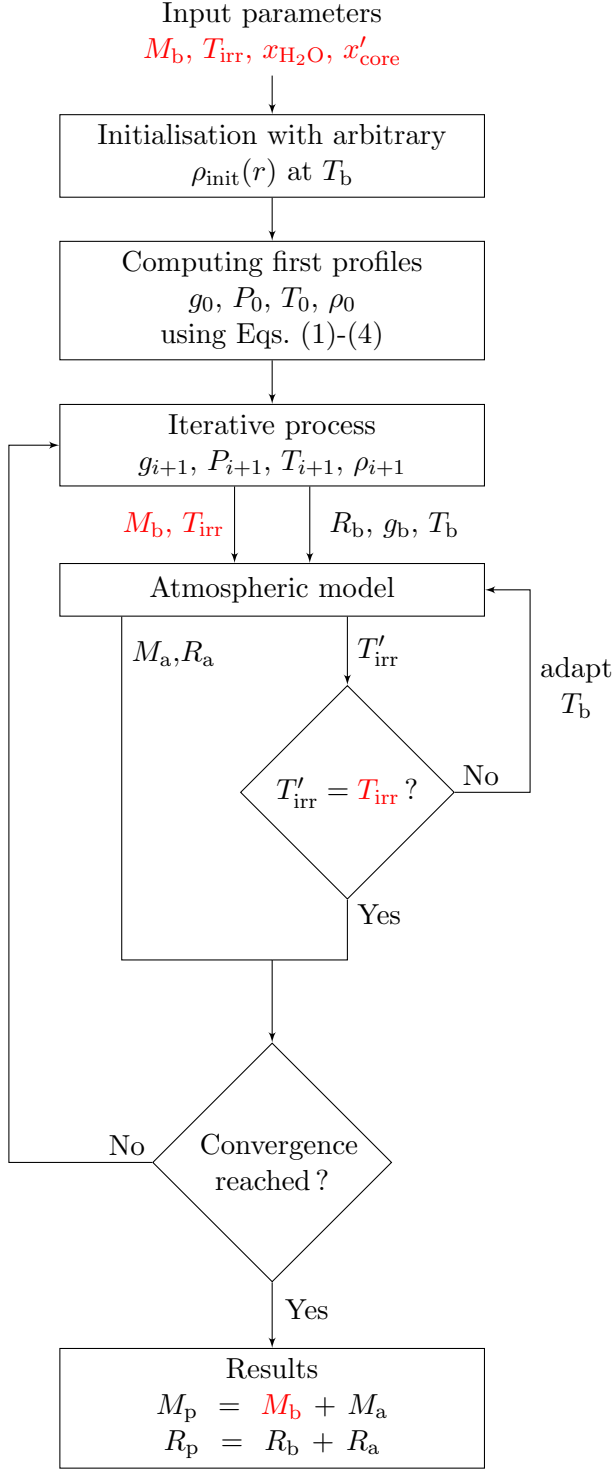}
	\caption{Numerical scheme used to produce mass-radius relationships. Quantities in red are fixed parameters that do not change throughout the computation.}
	\label{fig:scheme}
\end{figure}

The iterative process at work in our interior model is the following:
\begin{itemize}
	\item First, an arbitrary density profile $\rho_\mathrm{init}$ is given;
	\item Equations (\ref{eq:gauss})--(\ref{eq:solve_eos}) produce gravity, pressure, temperature and density profiles, in that order, at each layer;
	\item the computed density profile $\rho_i$ is used to compute the next iteration $\rho_{i+1}$ until convergence is reached.
\end{itemize}
Apart from the compositional and thermodynamic parametrizations, the model takes as inputs the mass within the boundary $M_\mathrm{b}$, the boundary pressure fixed to $P_\mathrm{b}=300$ bar and the boundary temperature $T_\mathrm{b}$, and produces the planet's radius at the boundary $R_\mathrm{b}$ (also giving $g_\mathrm{b}$). The atmospheric model takes as inputs the planet's boundary conditions ($M_\mathrm{b}$, $g_\mathrm{b}$, $T_\mathrm{b}$ and $P_\mathrm{b}=300$ bar) and gives the atmosphere's mass, thickness (at 0.1 Pa) and irradiation temperature $T_\mathrm{irr}^\prime$. To connect the two models, we implemented the atmospheric data grid with trilinear interpolation directly inside the MSEI model, which require a second iteration process that finds $T_\mathrm{b}$ that matches the input $T_\mathrm{irr}$. The corresponding numerical scheme is given in Fig. \ref{fig:scheme}.

\section{Trilinear interpolation} \label{sec:trilinear}

Considering a data grid that gives values of a function $f(x)$ at specific points $x$, linear interpolation is a method that allows to estimate values of $f$ between two points $x_a$ and $x_b$ by assuming $f$ is linear, 
giving the formula:

\begin{eqnarray}
	f(x) = f(x_a) +  (x-x_a)\frac{f(x_b)-f(x_a)}{x_b-x_a} = f(x_b)\frac{x-x_a}{x_b-x_a} + f(x_a)\frac{x_b-x}{x_b-x_a}.
\end{eqnarray}

The right hand side can be seen the opposite-length-weight average of the closest available data points (value $f(x_b)$ has the weight of the length from $x_a$, and value $f(x_a)$ has the weight of the length from $x_b$, 
hence ``opposite"). The concept of weight average is especialy useful as we can generalize this method to D-linear interpolation. Consider a D-dimensional box (or hyperrectangle or D-orthotope) the $2^D$ vertices of which have coordinates $\boldsymbol{x}$, a D-dimensional vector, and values of the function $f$ at each vertex are known. The value of $f(\boldsymbol{x})$ within the box can be estimated by taking the average of $f(\boldsymbol{x})$ at vertices, weighted by the opposite vertex D-volume.

For a bilinear interpolation, the 2-volume is a surface. For trilinear interpolation, the 3-volume of a physical function $f(x,y,z)$ has actually the unit of $x\cdot y \cdot z$. In our case, the atmospheric model of \cite{Mar19} gives OLR, $A$ (Albedo), $M_\mathrm{a}$ (computed by integrating the $\rho(z)$ profile) and $R_\mathrm{a}$ as a function of $M_\mathrm{b}$, $g_\mathrm{b}$ and $T_\mathrm{b}$.

Mathematicaly, D-linear interpolation has two main flaws and two limitation:
\begin{itemize}
	\item the derivative is poorly estimated within the box, and the interpolated function is not differentiable at facets;
	\item the method being an averaging of the closest vertices, it will be of limited use for rapidly varying functions;
	\item values of $f$ must exist at all vertices, if one or more are unavailable, the interpolation fails;
	\item all facets of the box must be orthogonal to each other (i.e. the box is defined by only two opposite vertices, the min and max value for each variable).
\end{itemize}

These limitations can be resolved by other types of interpolation, or more efficiently by fit of a function based on physical arguments as was cleverly done in \cite{Tu20}. In our case, values produced by the model are evolving smoothly and with regular tendencies. The strength of D-linear interpolation is that at a specific node of the data grid, the D-linear interpolation gives exactly values of this node.

Note that extrapolation outside data range is possible. We allow our model to extrapolate beyond the available grid, but these cases are marked as "extrapolated", and assumed incorrect. In the worst case, the extrapolation can return an albedo greater than 1, which would result in an imaginary irradiation temperature according to Eq. (\ref{eq:teq}).


\end{document}